\documentclass[%
aip,
jcp,
amsmath,amssymb,
reprint,
]{revtex4-2}

\usepackage{graphicx}
\usepackage{dcolumn}
\usepackage{bm}

\usepackage[utf8]{inputenc}
\usepackage[T1]{fontenc}
\usepackage{mathptmx}
\usepackage{etoolbox}
\usepackage{xr}

\usepackage{xcolor}

\makeatletter
\def\@email#1#2{%
	\endgroup
	\patchcmd{\titleblock@produce}
	{\frontmatter@RRAPformat}
	{\frontmatter@RRAPformat{\produce@RRAP{*#1\href{mailto:#2}{#2}}}\frontmatter@RRAPformat}
	{}{}
}%
\makeatother
\begin{document}
	
	\preprint{AIP/123-QED}
	
	\title[]{Ultrafast internal conversion and photochromism in gas-phase salicylideneaniline}
	\author{Myles C. Silfies}
	\affiliation{Department of Physics and Astronomy, Stony Brook University, Stony Brook, New York 11794, USA\looseness=-1}
	\author{Arshad Mehmood}
	\affiliation{Department of Chemistry, Stony Brook University, Stony Brook, New York 11794, USA}
	\affiliation{Institute for Advanced Computational Science, Stony Brook University, Stony Brook, New York 11794, USA\looseness=-1}
	\author{Grzegorz Kowzan}
	\affiliation{Department of Chemistry, Stony Brook University, Stony Brook, New York 11794, USA}
	\affiliation{Institute of Physics, Faculty of Physics, Astronomy and Informatics, Nicolaus Copernicus University in Toru\'{n}, ul. Grudziadzka 5, 87-100 Toru\'{n}, Poland}
	\author{Edward G. Hohenstein}
	\affiliation{QC Ware Corporation, Palo Alto, California 94301, USA}
	\author{Benjamin G. Levine}
	\affiliation{Department of Chemistry, Stony Brook University, Stony Brook, New York 11794, USA}
	\affiliation{Institute for Advanced Computational Science, Stony Brook University, Stony Brook, New York 11794, USA\looseness=-1}
	\author{Thomas K. Allison}
	\affiliation{Department of Physics and Astronomy, Stony Brook University, Stony Brook, New York 11794, USA\looseness=-1}
	\affiliation{Department of Chemistry, Stony Brook University, Stony Brook, New York 11794, USA}
	\email{thomas.allison@stonybrook.edu}
	\date{\today}

\begin{abstract} 
Salicylidenaniline (SA) is an archetypal system for excited-state intramolecular proton transfer (ESIPT) in non-planar systems.
Multiple channels for relaxation involving both the keto and enol forms have been proposed after excitation to S$_1$ with near-UV light.
Here we present transient absorption measurements of hot gas-phase SA, jet-cooled SA, and SA in Ar clusters using cavity-enhanced transient absorption spectroscopy (CE-TAS). 
Assignment of the spectra is aided by simulated TAS spectra, computed by applying time-dependent complete active space configuration interaction (TD-CASCI) to structures drawn from nonadiabatic molecular dynamics simulations.
We find prompt ESIPT in all conditions followed by the rapid parallel generation of the trans-keto metastable photochrome state and fluorescent keto state in parallel.
Increasing the internal energy increases the photochrome yield and decreases the fluorescent yield and fluorescent state lifetime observed in TAS.
In Ar clusters, internal conversion of SA is severely hindered but the photochrome yield is unchanged.
Taken together, these results are consistent with the photochrome being produced via the vibrationally excited keto population after ESIPT. 
\end{abstract}


\maketitle 

\section{Introduction}
\begin{figure*}
	\includegraphics[width=\linewidth]{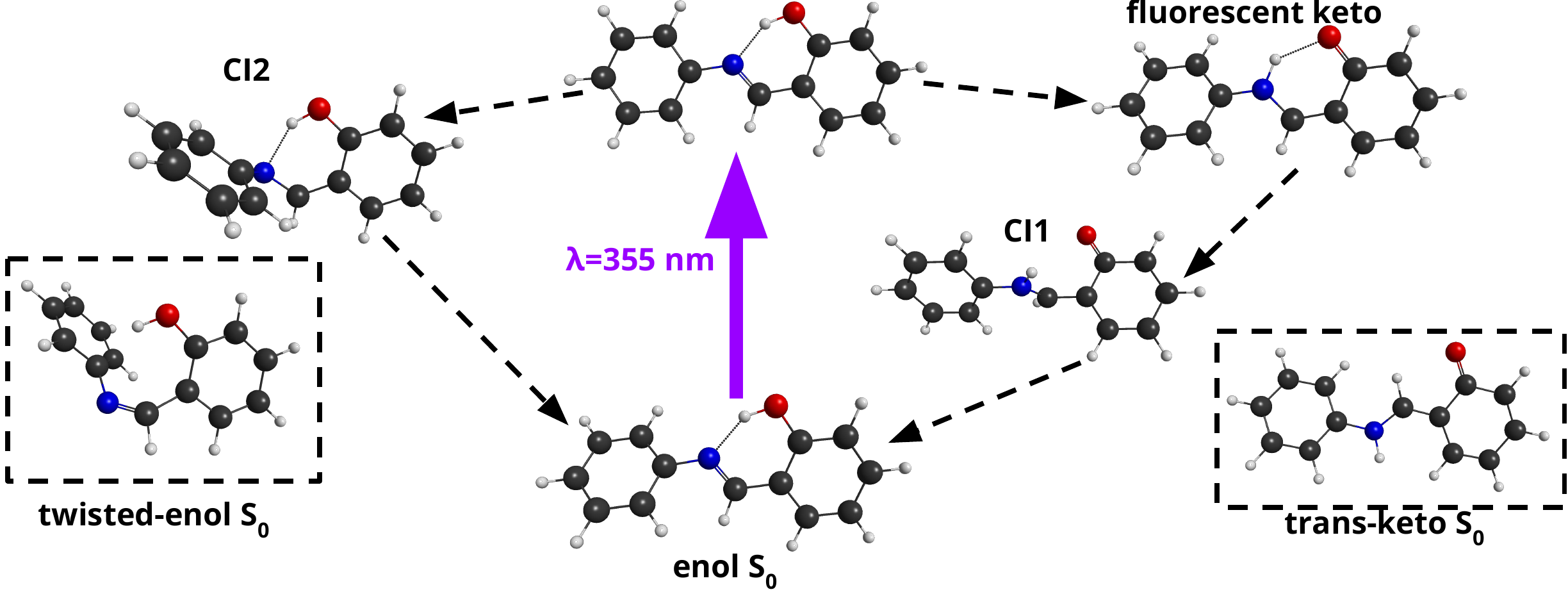}
	\caption{General overview of relaxation schemes for SA excited to S$_1$. After excitation, the excited enol can either (right side) undergo ESIPT and internally convert along the keto pathway through CI1 or (left side) remain in the enol tautomer and relax via CI2 by rotating about the central C-N bond. Bottom left and right insets are alternative ground state minima that are potential photochrome geometries which may be reached via the enol or keto relaxation pathways, repsectively.}
	\label{fig:scheme}
\end{figure*}

Tracking and understanding the redistribution of internal energy has been one of the main goals of ultrafast spectroscopy since its inception.
In photochromic molecules, the absorption of light causes large electronic reorganization leading to a reversible change in absorption and a corresponding material color change.\cite{uchida_AngewChemIntEd2004}
Often, these changes are driven by excited state intramolecular proton transfer (ESIPT) reactions and/or rapid isomerization on sub-picosecond timescales, requiring ultrafast techniques to follow the dynamics.
Additionally in the case of ESIPT, the redistribution of energy results in a large ($\sim 1$ eV) Stokes shifted fluorescence.\cite{joshi_Molecules2021,zhou_AccChemRes2018}
Recently, there has been a significant amount of work to use photochromic and/or ESIPT molecules in applications such as textiles,\cite{periyasamy_TextProg2017} optical memories,\cite{lim_JAmChemSoc2006,irie_ChemRev2000} and sensors.\cite{chen_NatCommun2017, mishra_JPhysChemA2004,qin_JMaterChemC2015}

Even as the absorption and emission shifts found in ESIPT/photochromic materials are now being exploited for consumer and industrial applications, many foundational spectroscopy studies continue on increasingly larger systems as well as in the closely related field of proton-coupled electron transfer (PCET).\cite{cukier_AnnuRevPhysChem1998,hammes-schiffer_ChemRev2010}
As the complexity of the molecule increases, so does the difficulty in understanding the relaxation dynamics and potential photochromism mediated by proton or hydrogen transfer as opposed to internal conversion mediated by other degrees of freedom. 
One of the most studied photochromic systems exhibiting competing dynamic pathways is the archetypal Schiff base salicylideneaniline (SA) depicted in Fig.\ \ref{fig:scheme}.
In solution and solid phases, SA becomes red after irradiation with UV light and remains trapped in this metastable state that can be reversed via irradiation by blue light. \cite{higelin_ChemicalPhysics1983,ottolenghi_TheJournalofChemicalPhysics1967,rosenfeld_PHOTOCHROMICANILSStructPHOTOISOMERSThermRelaxProcess1973,kownacki_ChemicalPhysicsLetters1994}

Although still under debate, the general photoinduced relaxation scheme was recently summarized and improved by Pijeau et al. with high-level dynamics simulations\cite{pijeau_JPhysChemA2018} which built off of earlier work. \cite{ortiz-sanchez_JChemPhys2008,zgierski_JChemPhys2000,sporkel_JPhysChemA2013}
This scheme, shown in Fig.\ \ref{fig:scheme} is described as follows.
In the ground state, SA is in the enol form and nonplanar with a $\approx 35^\circ-50^\circ$ twist around the anilino C-N bond depending on the level of theory or experimental data.\cite{destro_ActaCrystB1978,pijeau_JPhysChemA2018,ortiz-sanchez_JChemPhys2008,zgierski_JChemPhys2000}
Theory indicates that ESIPT is more likely to occur from a planar structure.\cite{zgierski_JChemPhys2000,ortiz-sanchez_JChemPhys2008,pijeau_JPhysChemA2018}
Upon UV excitation to the S$_1 (\pi,\pi*)$ state ($\lambda_{\mathrm{max}}\approx 350 \mathrm{ nm}$), SA must first planarize before undergoing ESIPT to form the excited keto form.
However, the ground state potential energy depends only weakly on the anilino C-N twist angle so even at low temperatures there is a nonzero probability of planar geometries.\cite{sekikawa_JPhysChemA2013} 

The keto form populated after ESIPT evolves either into the photochromic state, a fluorescent keto form, or undergoes internal conversion back to the ground state via a conical intersection (CI) followed by ground-state back proton transfer.
The keto CI, labeled ``CI1'' on Fig.\ \ref{fig:scheme}, was found theoretically and requires a rotation of nearly $90^\circ$ around the phenolic C-C bond. \cite{pijeau_JPhysChemA2018}
Using DFT calculations, Ortiz-S\'{a}nchez et al. found the ground state keto minimum structure shown on the bottom right of Fig.\ \ref{fig:scheme}, which has a full $180^\circ$ rotation about the same phenolic bond, and this is widely accepted as the photochrome geometry.\cite{yuzawa_ChemicalPhysicsLetters1993,avadanei_JPhysChemA2015,ortiz-sanchez_JChemPhys2008,nakagaki_BCSJ1977}
The sequencing and branching ratios of these excited keto decay pathways is not well understood or agreed upon.

Additionally, due to the nonplanar structure, a competing relaxation channel involving the excited enol without ESIPT has been proposed. \cite{pijeau_JPhysChemA2018,ortiz-sanchez_JChemPhys2008,zgierski_JChemPhys2000}
This pathway, shown on the left side of Fig.\ \ref{fig:scheme}, requires a large internal twist of $90^\circ$ around the central C-N bond to reach CI2 and relax to the ground state.
A secondary enol ground state minimum was found along this trajectory and is shown in the bottom left of Fig.\ \ref{fig:scheme}. 
This twisted enol geometry is also a candidate for the photochromic state, especially if excited with higher energy light.\cite{sliwa_PhotochemPhotobiolSci2010,avadanei_JPhysChemA2015,ortiz-sanchez_JChemPhys2008}
Pijeau et al. investigated the branching ratios and internal conversion rates of the 2 main channels and found that 80\% of the excited-state population relaxes via CI1 after ESIPT within 800 fs and 20\% via the enol CN twist channel in $\approx$ 250 fs \cite{pijeau_JPhysChemA2018}. 

SA has been studied using a variety of ultrafast techniques to observe the competing reaction dynamics and determine the nature of the various decay channels. 
Mitra and Tamai recorded the first femtosecond transient absorption spectra (TAS) in various solvents and found an instrument-limited proton transfer time and a monoexponential decay of both excited-state absorption (ESA) and stimulated emission (SE) with 4 ps time constant in cyclohexane, as well as a long-lived signal assigned to the photochrome.\cite{mitra_ChemicalPhysicsLetters1998,mitra_PhysChemChemPhys2003}
They proposed, like many of the early studies,\cite{kownacki_ChemicalPhysicsLetters1994,barbara_JAmChemSoc1980} that the initially hot keto state relaxes into either a relaxed fluorescing keto or the photochrome within the first several hundred fs and then the fluorescent state undergoes internal conversion on the picosecond time scale, with the lifetime depending on solvent environment.
TAS measurements by Zi\'{o}\l{}ek et al. further refined the proton transfer time to $< 50$ fs and an excited state relaxation time constant of 7 ps in acetonitrile.\cite{ziolek_PhysChemChemPhys2004}
Their interpretation finds a sequential model along the keto channel with the photochrome evolving from the relaxed keto state in competition with back-proton transfer on the ground state, disagreeing with previous work.
Fluorescence upconversion measurements in acetonitrile also agree with this model. \cite{rodriguez-cordoba_JPhysChemA2007}

In the gas phase, Sekikawa et al. performed time-resolved photoelectron spectroscopy (TRPES) in a helium seeded molecular beam. \cite{sekikawa_JPhysChemA2013}
Their results agreed with the solution TAS work on the ESIPT timescale being $\lesssim 50$ fs but found the intermediate decay time constant was 1.2 ps which was assigned to a combination of the ESIPT and CN twist channels. 
They also observed a long-lived feature which they assigned as the trans keto photochrome but suggested that the isomerization occured in the excited state. 
To further understand the dynamics, tunable pump measurements were performed which will be discussed in section \ref{sec:discuss} in comparison to our results. 
In general, it is difficult to determine if the lack of agreement between TAS and TRPES is due to the observable (optical absorption vs.\ photoionization) or the environment (solution vs.\ gas phase), and this is particularly true for a molecule with many competing relaxation channels, such as SA.

In this article we address the dynamics of salicylideneaniline after excitation to S$_1$ with $\lambda = 355$ nm using a combination of cavity-enhanced transient absorption spectroscopy (CE-TAS) and quantum chemistry/molecular dynamics calculations that directly simulate the observable.
CE-TAS\cite{silfies_PhysChemChemPhys2021, reber_Optica2016} uses a combination of high-power fibers lasers \cite{li_ReviewofScientificInstruments2016} and tunable cavity-enhanced frequency combs \cite{silfies_OptLett2020, chen_ApplPhysB2019} for broadband transient absorption measurements in dilute molecular beams.
CE-TAS acts as a halfway point between TAS and TRPES -- sharing the observable of the former and the environment(s) of the latter.
By varying the molecular beam conditions, we record dynamics of vibrationally hot SA (420 K), jet-cooled SA, and SA embedded in Ar clusters.
We directly compare the CE-TAS measurements to calculations of the transient absorption spectra using a newly-developed real time time-dependent complete active space configuration interaction (TD-CASCI) technique.\cite{mehmood_inprogress2023,peng_JChemTheoryComput2018} 

Overall, we find stronger agreement with the results of previous solution-phase TAS studies than the conclusions of gas-phase studies base on photoionization. 
We find prompt ESIPT after photoexcitation and relaxation dynamics in accordance with the previously proposed keto channels under all our studied conditions.
We do not observe any evidence of the CN twist channel within our detection window, but cannot rule it out completely as we calculate that the TAS signal for this channel should lie to the blue of our shortest 450 nm wavelength. 
Most strikingly, we find that for SA in Ar clusters, internal conversion of the fluorescent state is shut off while the photochrome yield remains unchanged. This supports a parallel mechanism for fluorescence and the photochrome state.
Comparing all the measurements and theory, we present a comprehensive picture of the keto relaxation channels of salicylideneaniline in section \ref{sec:discuss}.

\section{Methods}
\subsection{Experimental}
SA was purchased from TCI chemicals (97\%) and used as recieved. 
For all experiments, the sample is placed in a stainless steel cell and heated to 125$^\circ$ C to increase the vapor pressure. 
Either He or Ar is used as carrier gas for a planar supersonic expansion from a 5 mm $\times$ 0.2 mm slit nozzle, as described previously. \cite{silfies_PhysChemChemPhys2021}
Unless otherwise specified, the He stagnation pressure is 0.1 bar and the Ar stagnation pressure is 1 bar.
For all measurements shown, the interaction region is 3 mm above the nozzle, well within the expansion's ``zone of silence.'' \cite{miller_Atomicandmolecularbeammethods1988}

The cavity-enhanced transient absorption spectrometer operating at 100 MHz repetition rate used for all experiments is described in detail in ref.\ \onlinecite{silfies_PhysChemChemPhys2021} and the supplementary material.
The pump wavelength is 355 nm and the cavity-enhanced probe wavelength is tunable from 450 to 700 nm.\cite{silfies_OptLett2020,chen_ApplPhysB2019}
The CE-TAS signal $\Delta S$ is constructed from subtraction of two signals from counter-propagating cavity-enhanced frequency combs delayed by $T_{\text{pr}} \approx 5$ ns, as described previously.\cite{reber_Optica2016, silfies_PhysChemChemPhys2021}
For pump/probe signals with much shorter lifetime than $1/f_{\text{rep}} = 10$ ns, the CE-TAS signal $\Delta S$ is the same as one would normally record in conventional transient absorption spectroscopy setups.
However, for longer-lived signals there are additional subtleties which must be accounted for and we discuss these in subsequent sections and the supplementary material. 

Unless otherwise stated, all data shown are magic angle signals constructed from signals recorded with pump and probe polarizations parallel ($\Delta S_{\parallel}$) and perpendicular ($\Delta S_{\perp}$) via $\Delta S_{MA} = (\Delta S_{\parallel} + 2\Delta S_{\perp})/3$.
All broadband spectra (e.g. Fig.\ \ref{fig:He_specDAS}) shown are built up piecewise from pump/probe scans at 10--12 discrete probe wavelengths, taking into account the wavelength-dependent cavity finesse.
To generate contour plots from the individual scans, a marching squares algorithm is used for interpolation. 
In the case of data taken with large amounts of Ar carrier gas, producing Ar clusters around the SA molecules that suppresses internal conversion, signal amplitudes are scaled by fluorescence signals recorded independently during the pump/probe measurements.\cite{silfies_PhysChemChemPhys2021}
For data taken with He carrier gas, the fluorescence signal was too weak for the current instrument sensitivity, so the data are unnormalized. 
However, when using He the molecule pickup is more stable than in the Ar case, most likely due to the lower stagnation pressures and the lack of clustering, and we find good reproducibility across multiple data sets even without fluorescence normalization.

Since the previously-measured ESIPT occurs below our instrument response of $\approx 200$ fs,\cite{ziolek_PhysChemChemPhys2004,sekikawa_JPhysChemA2013} fitting the rising edge of the signal to an error function is used for aligning time zero for each scan in the spectrum. 
We verified that the rising edge of the signal is indeed instrument-response-limited by separately measuring the the instrument response function at several probe wavelengths using 2-photon absorption in gas-phase carbon disulfide immediately following an SA measurement (supplementary material Fig.\ \ref{supp-fig:int_IRF}).

\subsection{Modeling}
\label{sec:modeling}
A standard tool for analyzing transient absorption spectra is global analysis (GA),\cite{vanstokkum_BiochimicaetBiophysicaActaBBA-Bioenergetics2004} in which the transient absorption signal is modeled as a sum of decay associated spectral (DAS) components $\mathrm{X}_n(\lambda)$ each with its own exponential decay with characteristic time constant, $\tau_n$, viz.
\begin{equation}
	G(\lambda,t) = \textrm{IRF}(t) \otimes \sum_\textrm{n} \textrm{X}_\textrm{n}(\lambda)\exp(-t /\tau_\textrm{n}) \label{eqn:normal_GA}
\end{equation}

In equation (\ref{eqn:normal_GA}) the GA model is convolved with a Gaussian instrument response function $\textrm{IRF}(t)$.
For many CE-TAS experiments on molecules with short-lived signals with $\tau \ll 1/f_{\text{rep}}$ this standard GA model is sufficient since the CE-TAS signal $\Delta S$ is essentially the same as TA signals recorded by conventional transient absorption spectrometers.
However, for data with long-lived components, with $\tau_n \gtrsim 1/f_{\text{rep}}$, it is more appropriate to use a modified model which accounts for multiple excitation of the sample and reference-pulse subtraction in the following manner
\begin{equation}
	\Delta S_{\text{model}} (\lambda,t) = \sum_{m=0}^{N} G(\lambda,t+m/f_{\text{rep}})-G(\lambda,t+ m/f_{\text{rep}} + T_{\text{pr}}) \label{eqn:CETAS_GA} \;.
\end{equation}
In equations (\ref{eqn:normal_GA}) and (\ref{eqn:CETAS_GA}), $G(\lambda,t)$ represents the intrinsic molecular dynamics and $\Delta S_{\text{model}}(\lambda,t)$ is the modeled CE-TAS signal. 
The time offset in the subtracted signal is due to the reference pulse reaching the sample $T_{\text{pr}}$ before the probe. 
$N$ is the approximate number of pump pulses that molecules see as they fly through the focus. 
For the modeling shown here, we use $N = 20$, but the DAS are not sensitive to this choice as long as it is much larger than 1.
Note that although the \emph{sample} is excited by $\sim N$ pulses, as discussed in ref. \onlinecite{silfies_PhysChemChemPhys2021}, the excitation density is sufficiently low that multiple excitation of the same \emph{molecule} is negligible.
All experimental data is modeled using equation (\ref{eqn:CETAS_GA}), although for the isolated molecule with only one long-lived component similar results are obtained using either (\ref{eqn:normal_GA}) or (\ref{eqn:CETAS_GA}).
Also note that for modeling CE-TAS data we use a wavelength dependent IRF.
The fit parameters are optimized by minimizing the reduced $\chi^2$ using a Levenberg-Marquardt global fitting algorithm.
A parallel model is used because the proposed relaxation channels occur in parallel following the initial excitation and not in sequence.
Additionally, any photochrome signature in the fit is sufficiently slow such that the DAS for the photochrome does not depend on whether a parallel or sequential model is used. 

\subsection{Theoretical Calculations}
The molecular dynamics calculations of Pijeau et al. \cite{pijeau_JPhysChemA2018} are used as the starting point for theoretical work.
These calcualtions use the \textit{ab-initio} multiple spawning (AIMS) method\cite{ben-nun_JPhysChemA2000,curchod_ChemRev2018} for modelling nonadiabatic molecular dynamics.  
The potential energy surfaces were computed on-the-fly via complete active space configuration interaction with density functional embedding correction. \cite{pijeau_JChemTheoryComput2017}  
The $\omega$PBEh functional \cite{rohrdanz_TheJournalofChemicalPhysics2009} and floating occupation molecular orbitals\cite{slavicek_JChemPhys2010,hohenstein_TheJournalofChemicalPhysics2016,hohenstein_TheJournalofChemicalPhysics2015,hollas_JChemTheoryComput2018} (FOMO) were used.
An active space of two electrons in two orbitals was employed.  
The FOMO temperature was set to 0.35 Hartree.  
We will abbreviate this method $\omega$PBEh-FOMO(0.35)-CAS(2,2)CI/6-31g**, going forward. 

The direct calculation of the spectroscopic observables from ab initio molecular dynamics simulations enables more definitive assignments of spectral features than is possible by the indirect comparison of experimentally observed lifetimes to simulations.\cite{hudock_JPhysChemA2007,borrego-varillas_NatCommun2021,chakraborty_JPhysChemLett2021,kochman_JPhysChemA2021,avagliano_JComputChem2022}
The TAS signal is computed theoretically from the AIMS simulations.
A detailed description of the method for calculation of the TAS spectrum is included in supplementary material section \ref{supp-sup:sec:sim}, so here we provide only a short overview.  
Representative geometries are systematically drawn from the AIMS simulations via a clustering algorithm.
Specifically, 80 distinct conformations are sampled from each $\approx$ 24 fs window of the dynamics simulations, for a total of 6720 structures.
The ESA and SE are computed via three time-dependent (TD-) CASCI method\cite{peng_JChemTheoryComput2018,durden_JChemTheoryComput2022} simulations at each geometry, with light polarized in the x, y, and z directions, respectively.
In doing, we take advantage of the fact that time-dependent electronic structure methods are a robust and efficient method to compute molecular absorption spectra.\cite{li_ChemRev2020,degiovannini_ChemPhysChem2013,bruner_JChemTheoryComput2016,nascimento_JChemTheoryComput2016,goings_WIREsComputMolSci2018}
A larger (8,8) active space is used to calculate the TAS spectrum and the FOMO temperature was reduced to 0.25 Hartree, with the other parameters of the electronic structure method as described above for the dynamics ($\omega$PBEh-FOMO(0.25)-TD-CAS(8,8)CI/6-31g**). 
The electronic wavefunctions at each geometry are excited by a $\delta$ function pulse and then propagated for 45 fs with a 0.003 fs steps.   
Fourier transforms of the resulting correlation functions provide spectra with a spectral resolution of 0.11 eV ($\Delta \lambda = 28$ nm at $\lambda = 450$ nm; $\Delta \lambda = 74$ nm at $\Delta \lambda = 700$ nm). 
The ESA and SE signals are separately shifted by +0.944 and -1.595 eV respectively to better agree with more trustworthy complete active space second order perturbation theory calculations of ESA and the experimental absorption maximum, respectively.
In total, the spectra presented required the simulation of 0.9 ns of electron dynamics, which was enabled by our previously-reported GPU-accelerated direct TD-CASCI algorithm.
All electronic structure calculations were carried out using the TeraChem software package\cite{ufimtsev_JChemTheoryComput2009,seritan_WIREsComputMolSci2021,fales_JChemTheoryComput2015}.  
Additional methodological details can be found in supplementary materials, and a separate paper focused on the theoretical method for computing TAS is forthcoming.\cite{mehmood_inprogress2023}

To facilitate comparison with the experiment, we process the theoretical TAS results in a manner analogous to how the experiment is performed and analyzed. 
First, the theoretical TAS data from the AIMS/TD-CASCI are sampled at points with 15 nm spacing with a 3 THz FWHM instrument response, which approximately corresponds to the intracavity bandwidth with which each experimental TA trace is recorded.
The sampled dataset is then further convolved with a 200 fs FWHM Gaussian in time.

\section{Results}
\subsection{The Isolated Molecule}
\begin{figure}
	\includegraphics[width=\linewidth]{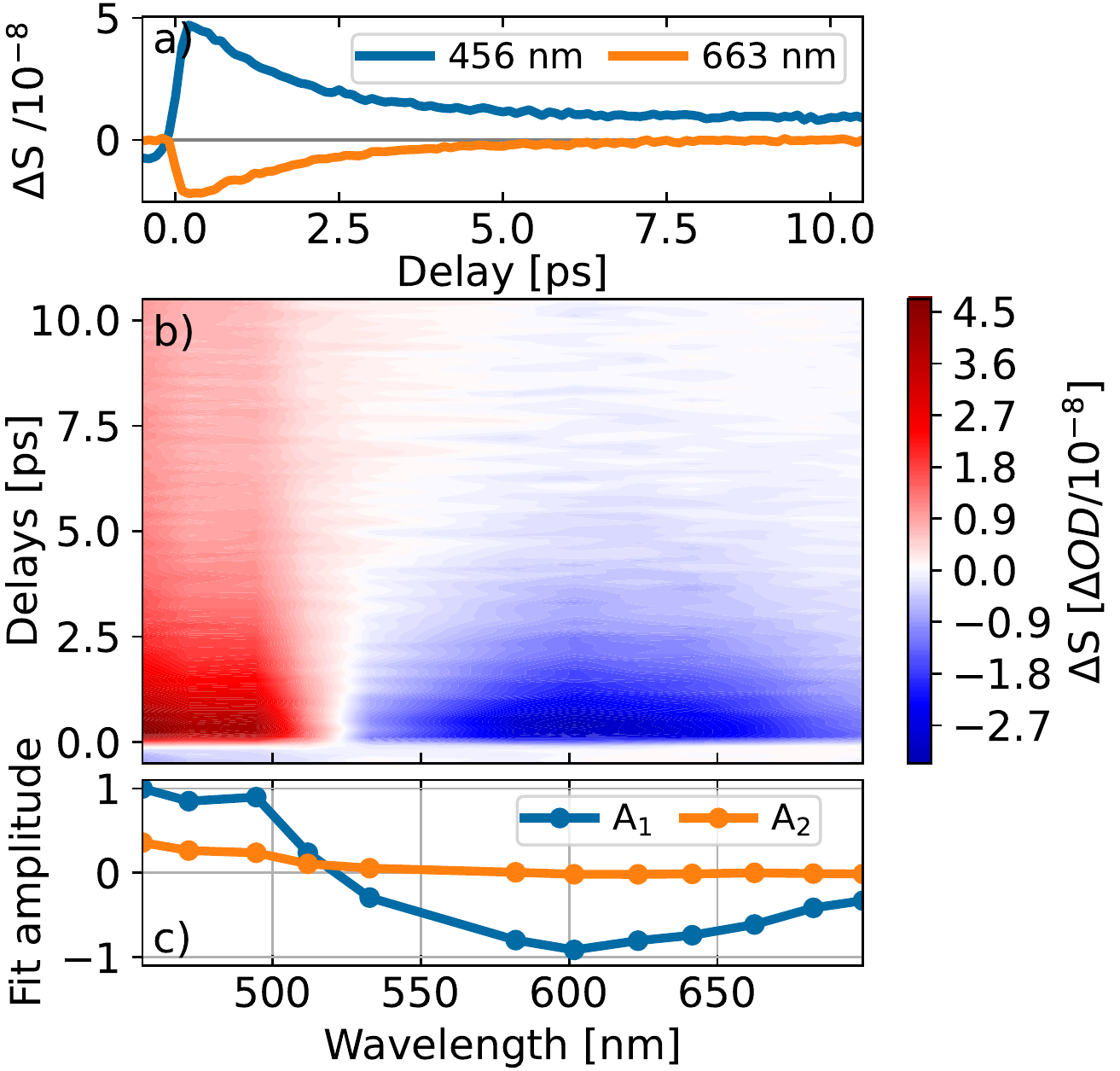}
	\caption{CE-TAS of isolated jet-cooled SA molecule. a) Raw signals at 2 representative probe wavelengths. b) TA spectrum constructed from the individual scans. c) DAS from global fit of b) with 2 components, normalized to the maximum of A$_1$. A$_1$ and A$_2$ are associated with time constants of 1.8 ps and >10 ns, respectively. Individual points are at the discrete probe wavelengths which are combined for the full spectrum. The origin and meaning of pre-time-zero signals is discussed in the text.}
	\label{fig:He_specDAS}
\end{figure}
The TA of SA cooled in He carrier gas is shown in Fig.\ \ref{fig:He_specDAS}.
Representative pump/probe scans are shown in Fig.\ \ref{fig:He_specDAS}a) and the full TA spectrum in Fig.\ \ref{fig:He_specDAS}b).
Note that the negative $\Delta S$ at negative time delays is due to long-lived excited-state absorption in the signal on the blue side of the spectrum. 
The origin of this pre time-zero signal is illustrated with examples in the supplementary material. 
This negative delay $\Delta S$ signal provides information on long-lived components of the true TA signals, and these long-lived DAS are extracted via modeling the full signal (i.e. both negative and positive delays) as discussed in section \ref{sec:modeling}. 

The SA ground-state minimum does not absorb in the visible range, therefore no bleach signal is considered,\cite{mitra_PhysChemChemPhys2003} i.e. all negative signals are from stimulated emission.
From modeling of the decay of the polarization anisotropy (i.e.\ the weighted difference between signals recorded with parallel and perpendicular relative polarizations) using the procedure of Felker et al.,\cite{felker_JChemPhys1987} we estimate the rotational temperature in the expansion to be 80 K, well above the condensation temperature of He, such that no clustering in the helium expansion is expected or observed.

We fit the data using equations (\ref{eqn:normal_GA}) and (\ref{eqn:CETAS_GA}) with $\mathrm{X}_n \equiv \mathrm{A}_n$.
The DAS for the isolated SA are shown in Fig.\ \ref{fig:He_specDAS}c) for GA with two time constants (reduced $\chi^2=1.9$). 
The associated time constants are $\tau_{\textrm{A}_1} = 1.8$ ps and $\tau_{\textrm{A}_2}$ >10 ns.
The points shown on the DAS are the discrete probe wavelengths that make up the spectrum.
The same fit procedure was also tested for 1 (reduced $\chi^2=51.5$) and 3 (reduced $\chi^2$=2.7) components. 

The short-lived $\mathrm{A}_1$ is composed of both negative and positive features in the spectrum indicating excited state absorption (ESA) and stimulated emission (SE) from the same state, and is the typical type of TAS signature seen for excited-state proton transfer in many molecules. 
The 1.8 ps time constant lies between those reported for gas-phase TRPES\cite{sekikawa_JPhysChemA2013} and solution-phase TAS.\cite{ziolek_PhysChemChemPhys2004, mitra_ChemicalPhysicsLetters1998,mitra_PhysChemChemPhys2003}
The long-lived $\mathrm{A}_2$ extends out from blue edge of the spectrum, is only positive, and remains nearly constant out to the maximum delay of the instrument (700 ps).
In the raw pump-probe traces, A$_2$ also appears as a negative signal for wavelengths less than 530 nm which is accounted for in the model as described in section \ref{sec:modeling}.
This feature is most likely ground-state absorption from the long-lived photochromic state observed in numerous previous experiments.
The assignment and origin of this state is further discussed in section \ref{sec:discuss}.

\begin{figure}
	\includegraphics[width=\linewidth]{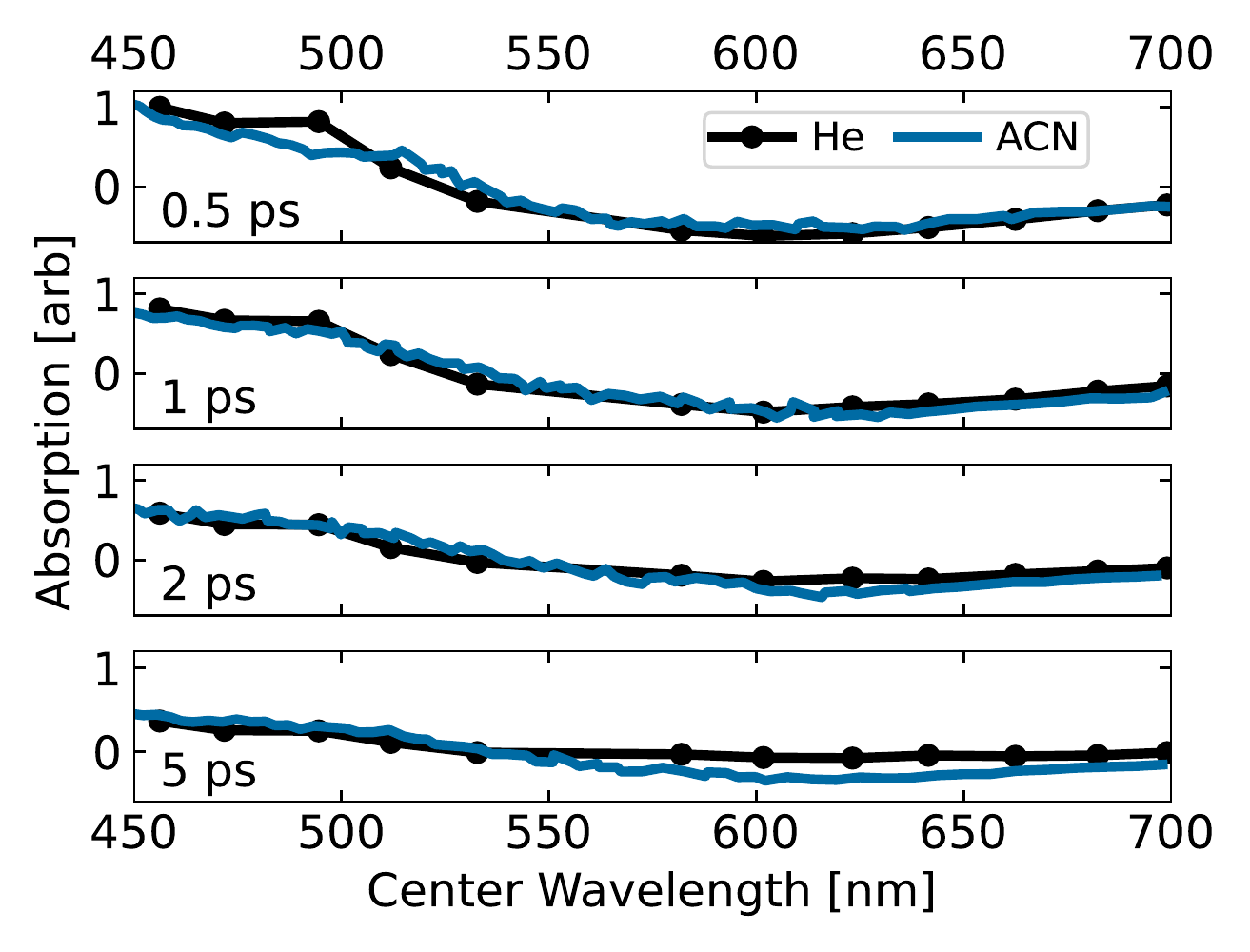}
	\caption{Lineout comparison between TAS of the jet-cooled SA molecule and in SA in acetonitrile (ACN) for different delay times. One global scaling factor is used between the two datasets at all times. ACN data from ref. \onlinecite{ziolek_PhysChemChemPhys2004}.}
	\label{fig:ACN}
\end{figure}

In Fig.\ \ref{fig:ACN}, we compare our TA spectra to those reported by Zi\'{o}\l{}ek et al.\cite{ziolek_PhysChemChemPhys2004} for SA excited at $\lambda = 390$ nm in acetonitrile. 
The data shown has only a single global scaling to overlap the bluest probe value at 0.5 ps and this scale factor is applied to all of the gas-phase data.
The spectra show remarkable similarity, without even a solvatochromic shift.
Additionally, Mitra and Tamai also measured solution-phase fs TAS of SA in several solvents and there are only minor spectral shifts between their data and our gas-phase measurements.\cite{mitra_ChemicalPhysicsLetters1998,mitra_PhysChemChemPhys2003}

While we do not vary the excitation energy in our experiments, we do vary the initial vibrational energy via coarse control of the temperature.
The nozzle is held at 155$^{\circ}$ C, and for an effusive beam (i.e. no carrier gas) or very low carrier gas stagnation pressures (quasi-effusive beam) such that there is no supersonic expansion, we expect the molecular temperature in the beam to be approximately the same as the nozzle.
In practice, we find the SA density in the beam to be more stable for quasi-effusive beam, and record ``hot SA'' data with 0.02 bar stagnation pressure of He. 
We have verified that identical dynamics are obtained with quasi-effusive beam and with no carrier gas and analyzing the rotational anisotropy, we find a rotational temperature for ``hot SA'' of 420 $\pm$ 20 K for the quasi-effusive beam, consistent with the nozzle temperature.

\begin{figure}
	\includegraphics[width=\linewidth]{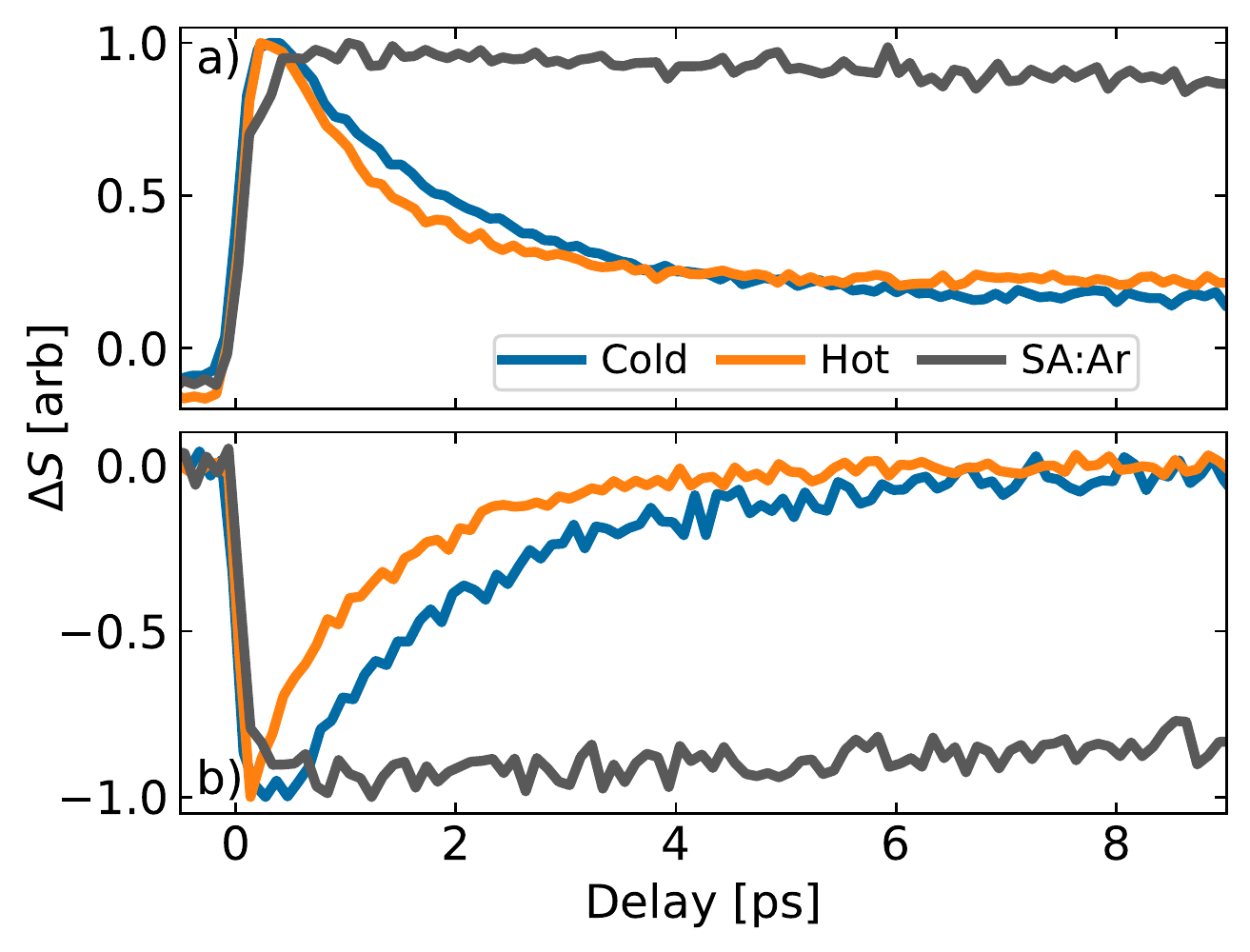}
	\caption{Comparison of SA dynamics under different molecular beam conditions. ``Cold'' data are taken with 0.1 bar He stagnation pressure, giving a supersonic expansion and jet-cooled molecules. ``Hot'' data are recorded from a quasi-effusive beam. The SA:Ar data are taken with 1 bar Ar stagnation pressure, forming large Ar clusters. a) $\lambda_{\text{probe}} = 490$ nm. b) $\lambda_{\text{probe}} = 616$ nm.}
	\label{fig:pressure}
\end{figure}

Fig.\ \ref{fig:pressure} compares pump/probe traces for ``hot SA'' with the quasi-effusive beam and ``cold SA'' recorded with a He stagnation pressure of 0.1 bar.
Higher He stagnation pressures do not alter the signal significantly.
The vibrationally hot molecule decays slightly faster and shows a slightly larger amplitude for the long-lived signal. 

\subsection{Salicylideneaniline in Argon Clusters}
\begin{figure}
	\includegraphics[width=\linewidth]{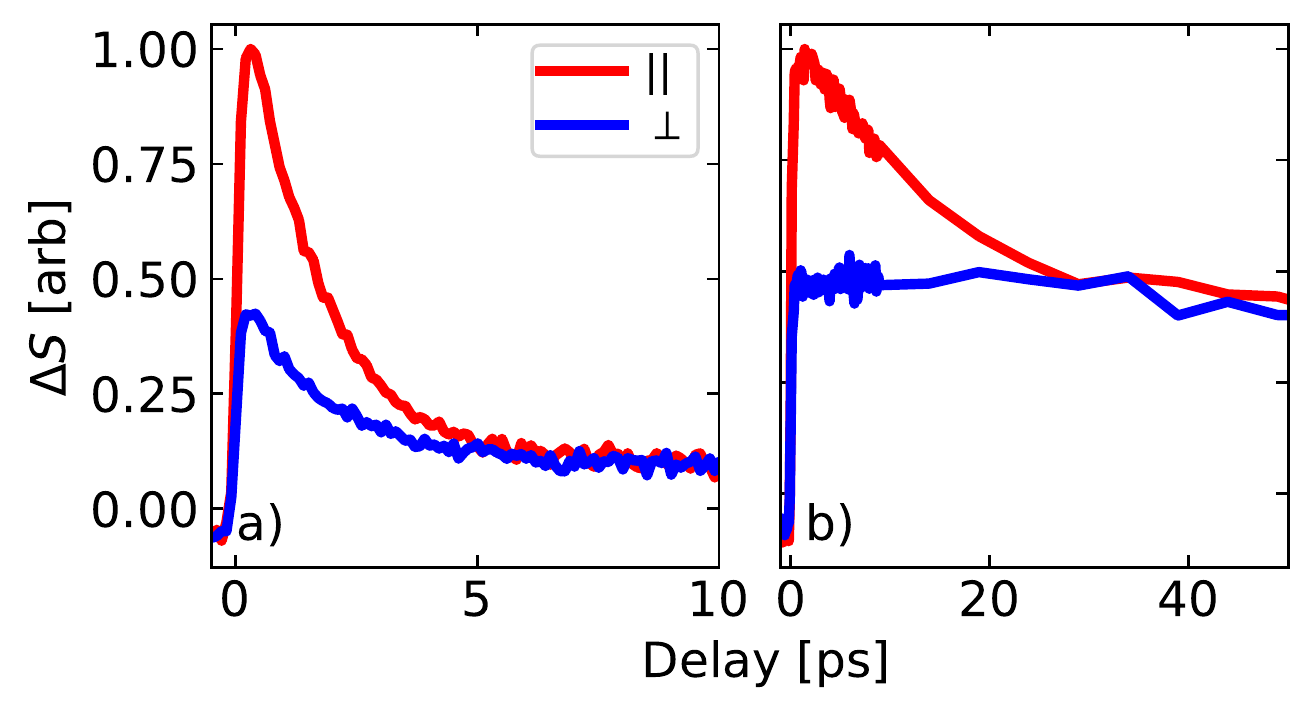}
	\caption{Rotational dynamics of isolated SA vs.\ SA:Ar. CE-TAS signals with pump and probe pulses parallel (red) and perpendicular (blue) for a) isolated jet-cooled SA and b) SA:Ar with 1 bar Ar stagnation pressure. $\lambda_{\text{probe}} = 490$ nm. Formation of large Ar clusters around the SA molecules dramatically slows rotational dephasing.}
	\label{fig:anisotropy}
\end{figure}
\begin{figure}
	\includegraphics[width=\linewidth]{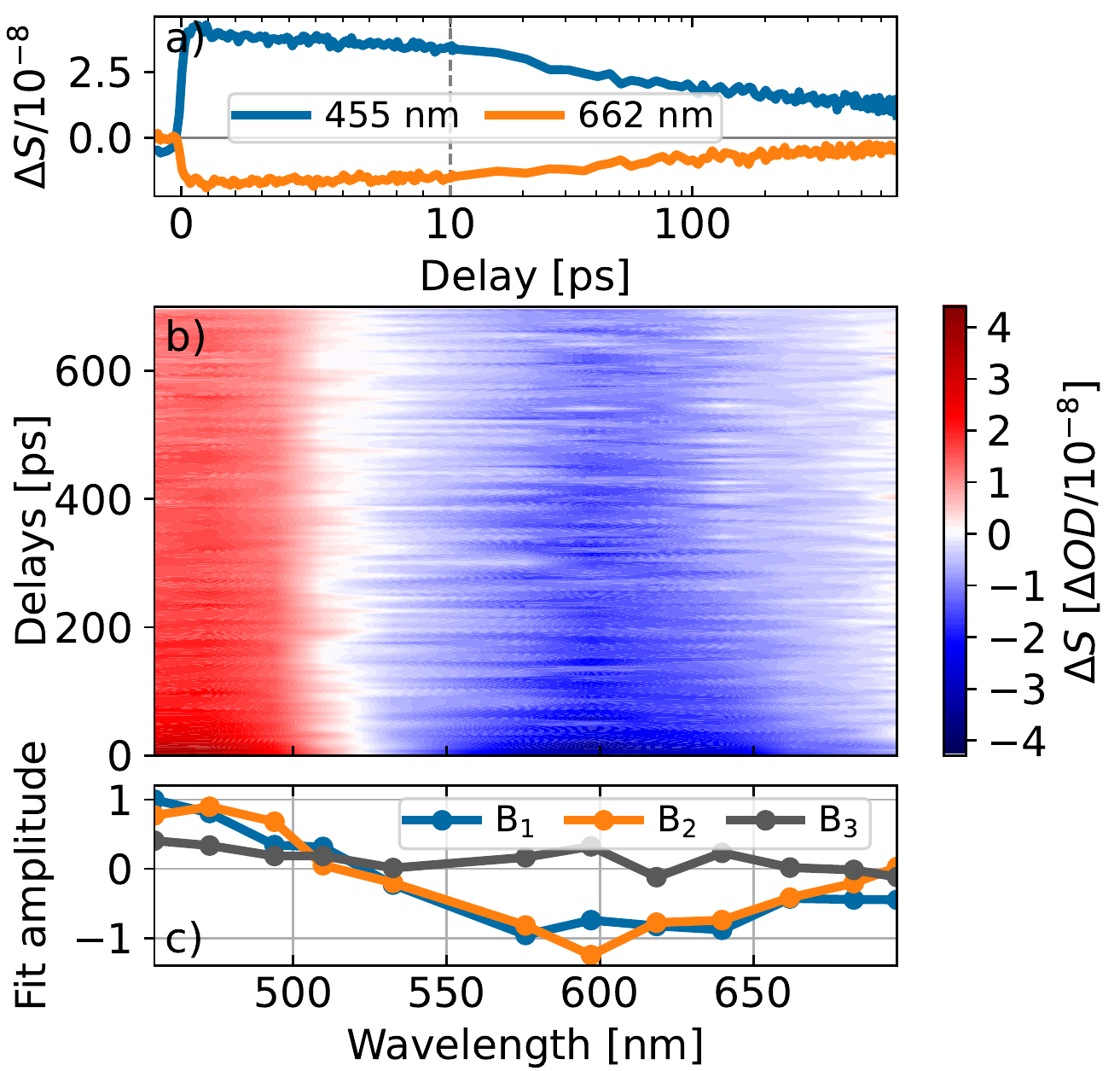}
	\caption{CE-TAS of SA:Ar from SA seeded in 1 bar of Ar. a) Pump/probe traces at 2 representative probe wavelengths. The time axis shown changes from linear to logarithmic at 10 ps to better show the signal at the negative time delays. b) TA spectrum constructed from the individual scans. c) DAS from global fit of b) with 3 components normalized to the maximum of B$_1$. $\tau_{\text{B}_1} = 24$ ps, $\tau_{\text{B}_2} = 850$ ps, and $\tau_{\text{B}_3} >10$ ns. Individual points are at the discrete probe wavelengths which were combined for the spectrum.}
	\label{fig:Ar_specDAS}
\end{figure}

To further understand the different parallel relaxation channels of SA, we perform experiments on SA in Ar clusters which we denote SA:Ar.
Similar to a rare-gas matrix environment, Ar clustering can affect the dynamics in two ways: (1) by providing a channel for the dissipation of vibrational energy and (2) by providing steric hindrance to large amplitude motions, such as the isomerizations shown in Fig.\ \ref{fig:scheme}. 
Figure~\ref{fig:pressure} compares the CE-TAS signals for the SA:Ar system to the isolated molecule. The SA:Ar signal decays much slower in the clustered sample than in the hot or cold SA case. 
From our fluorescence measurements, we can place a lower bound on the enhancement of the total fluorescence yield of SA:Ar vs.\ isolated SA of 100x.
Fluorescence enhancement in SA has also been previously reported in matrix isolation studies.\cite{barbara_JAmChemSoc1980}
The data in Fig.\ \ref{fig:pressure} were recorded at 1 bar since this provided a workable, stable signal for recording a full spectrum. 
Increasing the Ar pressure continues to enhance this effect until the pump/probe signal is nearly constant out to 700 ps at a stagnation pressure of 2.5 bar. 

We estimate the average cluster size by analyzing the rotational anisotropy.
Fig.\ \ref{fig:anisotropy} compares $\Delta S_{\parallel}$, $\Delta S_{\perp}$ data taken in a He expansion to an Ar expansion with 1 bar stagnation pressure.
The rotational anisotropy persists roughly 5 times longer due to the increased moment of inertia of the SA:Ar system.
Considering that the rotational dephasing time scales as $1/\sqrt{B}$,\cite{felker_JChemPhys1987} where $B$ is the rotational constant, we estimate an average number of 15 Ar atoms clustered to the SA molecules.\cite{buck_TheJournalofChemicalPhysics1996}

A full CE-TAS dataset for the SA:Ar system is shown in Fig.\ \ref{fig:Ar_specDAS} with representative scans shown in Fig.\ \ref{fig:Ar_specDAS}a) and the full spectrum in Fig.\ \ref{fig:Ar_specDAS}b).
Just as in the case of the individual lineouts discussed above, the entire spectrum decays with a significantly longer time constant than in the cold SA case above.
Despite the slower decay, the initial rise time remains instrument response limited, indicating rapid ESIPT rates unaffected by the cluster environment.

We fit the data using equations \ref{eqn:normal_GA} and \ref{eqn:CETAS_GA} with $\mathrm{X}_n \equiv \mathrm{B}_n$.
The DAS from the fit are shown in Fig.\ \ref{fig:Ar_specDAS}c) for 3 components (reduced $\chi^2$=4.2).
The associated time constants are $\tau_{\textrm{B}_1} = 32$ ps, $\tau_{\textrm{B}_2} = 1030$ ps, and $\tau_{\textrm{B}_3}$ >10 ns.
The same fit procedure was also tested for 2 (reduced $\chi^2=5.0$) and 4 (reduced $\chi^2$=4.1) time constants.
The 4-component fit is shown in supplementary material (Fig.\ \ref{supp-fig:Ar_ds_4}). 
The additional DAS for the 4-component fit has only a small amplitude at all wavelengths and a long decay time not apparent in the raw data, and thus was determined to be not significant. 
In general, the fits on the SA:Ar data are worse than the He case due to increased noise from turbulence from the higher pressure gas and scatter from bare Ar clusters.  

The overall shapes of the DAS are nearly identical to the He case in Fig.\ \ref{fig:He_specDAS}b), indicating that the clustering is not modifying the electronic energies significantly.
The only major change in the fit results is the additional component B$_1$.
B$_1$ and B$_2$ are nearly identical to each other and to A$_1$ which most likely indicates that the monoexponential A$_1$ feature in cold SA becomes a biexponential B$_1 \rightarrow $ B$_2$ decay in SA:Ar.
Previous fluorescence measurements in matrix isolation also observed bioexponential decay in the fluorescence.\cite{barbara_JAmChemSoc1980}

\subsection{Simulated spectrum}
\label{sec:res_th}
\begin{figure*}
	\includegraphics[width=\linewidth]{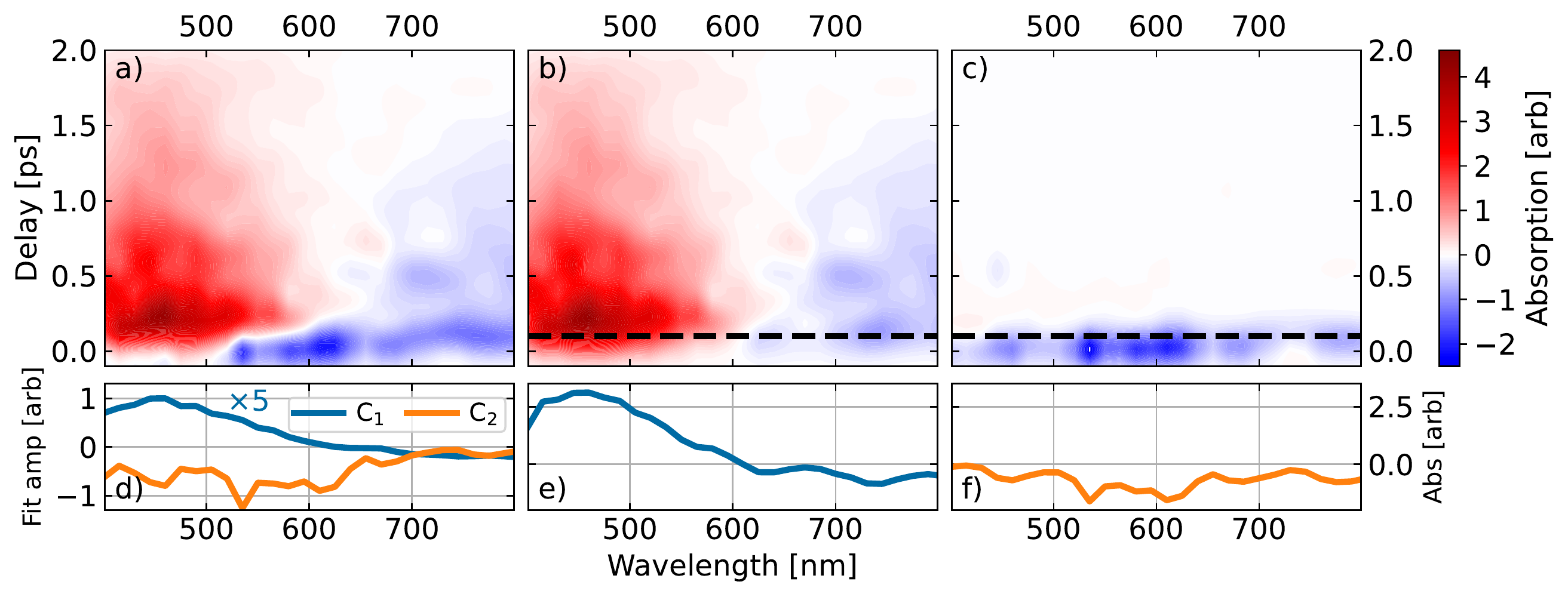}
	\caption{Theoretical TAS results from post-processed AIMS calculations containing a) all geometries, b) only keto geometries, and c) only enol geometries. d) DAS from a global fit of the spectrum in a). e) and f) show spectral lineouts using the right axis at 100 fs from the corresponding keto and enol spectra, respectively. See text for details regarding resampling and convolution with experimental resolution.}
	\label{fig:sim_specDAS}
\end{figure*}

The simulated TAS spectrum is presented in Fig. \ref{fig:sim_specDAS}a).  
The raw theoretical data is presented in supplementary material Fig.\ \ref{supp-fig:noconv}.  
Just as for the experimental data, a global fit was applied to the theoretical data, with $\mathrm{X}_n \equiv \mathrm{C}_n$ in equation (\ref{eqn:normal_GA}).
Fig.\ \ref{fig:sim_specDAS}d) shows the DAS from a global fit to the data with 2 components.
The associated time constants are $\tau_{\textrm{C}_1} = 0.62$ ps and $\tau_{\textrm{C}_2} = 20$ fs.
Because there is no meaningful error on each simulated point, the reduced $\chi^2$ is not useful, but removing or adding components resulted in either a poor or non-physical (i.e. two nearly identical time constants) fit.
The trajectories in the AIMS simulations performed by Pijeau et al. were ended once the initial S$_1$ wavepacket population decayed below 0.01, so there is no contribution of trajectories on the ground-state surface to the spectrum.\cite{pijeau_JPhysChemA2018} 
Importantly, this means that the photochrome signal is not present in the simulated TAS dataset but we address this drawback with static calculations discussed below.

A major benefit of directly computing experimental observables is that it is possible to analyze the spectrum be decomposing it into contributions that correspond to different molecular structures.  
To analyze the signature of proton transfer, we separate the spectrum into contributions from only keto or enol geometries by comparing O-H and N-H distances along the proton transfer coordinate.
When $d_{\textrm{O-H}} - d_{\textrm{N-H}}$ is positive (negative), these geometries are labeled enol (keto).
The spectral contributions from the keto and enol tautomers are shown in figures \ref{fig:sim_specDAS}b) and 7c), respectively.
For comparison with the DAS of Fig. \ref{fig:sim_specDAS}d), we plot lineouts of the spectra at 100 fs in Fig. \ref{fig:sim_specDAS}e) and \ref{fig:sim_specDAS}f).
From this comparison, it is clear that the C$_1$ DAS corresponds to the keto spectrum after ESIPT.

C$_2$ on figures \ref{fig:sim_specDAS}d) and \ref{fig:sim_specDAS}f) is entirely negative across most of the spectrum representing prompt, broadband stimulated emission from enol tautomers during only the first $\approx 100$ fs. 
The C$_2$ component corresponds well with the prompt enol geometry signal seen in Fig. \ref{fig:sim_specDAS}f), corresponding to near-planar enol geometries before and during ESIPT.

To further understand the spectral contribution from the CN twist relaxation mechanism, we filter the enol-only spectrum by CN twist angle, shown in supplementary material Fig.\ \ref{supp-fig:twist} for signals from geometries with angles less than $140^\circ$. 
In the ground state, this angle is $180^\circ$ for a planar orientation and $\approx 90^\circ$ near CI2 \cite{ortiz-sanchez_JChemPhys2008,pijeau_JPhysChemA2018} so any features on supplementary material Fig.\ \ref{supp-fig:twist} should be signatures of the internal conversion towards CI2.
Within the experimental observation window of 450-700 nm, all absorption or emission features from twisted enol geometries are more than an order of magnitude smaller than the main spectral components in Fig.\ \ref{fig:sim_specDAS}a).  
Thus, the CN twist channel may be present, but it is unlikely discernible in the TAS signal. 

Separate from the AIMS/TD-CASCI calculations we perform several TD-CASCI calculations at fixed geometries for the ground state local minima found by Ortiz-S\'{a}nchez et al. \cite{ortiz-sanchez_JChemPhys2008}. 
The calculated absorption spectra for both prospective photochrome conformers are included in supplementary material Fig.\ \ref{supp-fig:ortiz_abs}.
Our calculated trans keto photochrome (shown in the bottom right box of Fig.\ \ref{fig:scheme}) spectrum has an absorption peak centered near 480 nm. 
We also calculate the absorption spectrum of the Ortiz-S\'{a}nchez's twisted enol minimum (shown in the bottom left box of Fig.\ \ref{fig:scheme}), and found absorbance at $\approx$ 300 nm which is beyond the range probed in this experiment.

\section{Discussion}
\label{sec:discuss}
\begin{figure}
	\includegraphics[width=\linewidth]{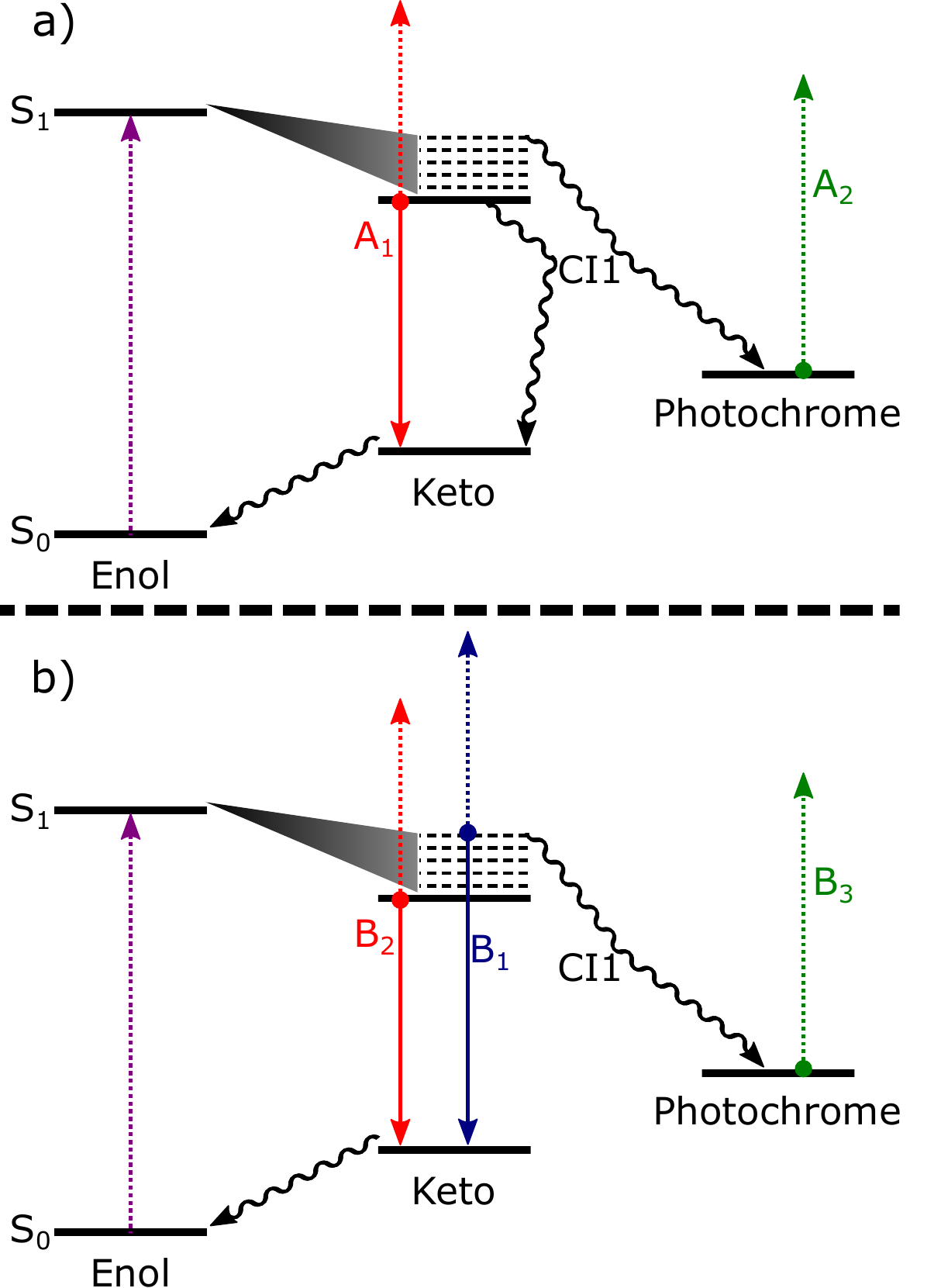}
	\caption{Proposed relaxation mechanism and photochrome pathway in a) isolated SA and b) SA:Ar. Dashed lines indicate absorption of light and solid lines are emission. Labels indicate DAS features from global analysis. In the isolated molecule, a), the initially excited enol undergoes ESIPT and generates the keto tautomer with a wide range of internal energies. The hottest of which immediately internally convert to CI1 and isomerize to the photochrome. The colder population in the relaxed fluorescent state also decays via CI1, but eventually returns to overall enol ground state. In SA:Ar, b), the the presences of the Ar cage slows down photochrome generation ($ \sim 0 \rightarrow$ 24 ps) and traps the vibrationally cold keto population such that it can only decay radiatively (1.8 ps $\rightarrow \; > 1 $ ns).}
	\label{fig:jablonski}
\end{figure}

Our overall proposed scheme for dynamics in both the isolated molecule and SA:Ar is shown in Fig.\ \ref{fig:jablonski}. 
Much of the dynamics is born out in the global analysis of the TA spectrum, which we describe step by step below.

In the isolated molecule, the initially excited enol quickly undergoes ESIPT, giving rise to the classic ESA/SE signature seen in the GA component A$_1$.
We assign A$_1$ to the fluorescent keto state, and this assignment is consistent with the theory component C$_1$ and the keto-filtered theory data of figure \ref{fig:sim_specDAS}b). 
This assignment is also in accord with previous work on SA and other ESIPT molecules.
At longer delays, after the fluorescent state has decayed, we assign the remaining long-lived signal, captured by A$_2$ to the keto photochrome.
This assignment is supported by the correspondence between A$_2$ and other reported spectra of the metastable photochrome state \cite{sliwa_PhotochemPhotobiolSci2010, ziolek_PhysChemChemPhys2004,mitra_PhysChemChemPhys2003,kownacki_ChemicalPhysicsLetters1994,ottolenghi_TheJournalofChemicalPhysics1967,rosenfeld_PHOTOCHROMICANILSStructPHOTOISOMERSThermRelaxProcess1973} and also our calculated absorption spectrum for the trans keto minimum.
We do not observe any signatures of the initially excited enol (C$_2$ and Fig.\ \ref{fig:sim_specDAS}c) in the experimental data, most likely due to our time resolution and excess pump energy, but we do note that it was observed in solution by fluorescence upconversion. \cite{rodriguez-cordoba_JPhysChemA2007} 

The 1.8 ps decay time of the excited keto state we observe is significantly slower than those observed in gas-phase TRPES by Sekikawa et al.\cite{sekikawa_JPhysChemA2013}
This highlights the role of the observable on the measured dynamics, since nominally these two measurements are taken under the same molecule and excitation conditions.
The shorter lifetime measured by TRPES could be due to energy windowing effects effects, \cite{barbatti_ChemicalPhysicsLetters2005,hudock_JPhysChemA2007,tao_JChemPhys2011} especially since the TRPES study was done using 1+2 photoionization.\cite{sekikawa_JPhysChemA2013}

Comparing to solution phase work, the gas-phase TA spectra very closely correspond to those measured in solution, as shown in the raw data in figure \ref{fig:ACN}.
This is not unexpected since the TA spectrum shape has been previously shown to be relatively insensitive to the choice of solvent. \cite{mitra_PhysChemChemPhys2003,mitra_ChemicalPhysicsLetters1998}
Regarding the dynamics, our 1.8 ps time constant is slightly shorter than the shortest S$_1$ keto lifetimes reported in solution phase work.
A range of solution-phase lifetimes measured via TAS between 3.5 and 50 ps have been previously reported, with a strong solvent dependence.\cite{ziolek_PhysChemChemPhys2004,mitra_PhysChemChemPhys2003,mitra_ChemicalPhysicsLetters1998}
The faster internal conversion in the absence of solvent supports the mechanism of internal conversion occurring at geometries very twisted compared to the ground state, as solvent can impede or slow down this large-amplitude motion.

Previous gas-phase experiments on SA varied the internal energy by changing the excitation wavelength.
Sekikawa et al.\ observed a larger long-lived photochrome signal as the excitation energy was increased and assigned this feature to enhanced ESIPT with larger excitation energy due to barriers to ESIPT at twisted geometries.\cite{sekikawa_JPhysChemA2013}
In our experiment, we vary the internal energy of the molecule via large changes in temperature.
Notably, we find that the appearance of the fluorescent keto signal (A$_1$) is equally prompt for both the hot SA and cold SA as shown in Fig. \ref{fig:pressure}, and thus find no evidence for a delay in the ESIPT at lower internal energies.
However, we do find that the photochrome yield is higher for the vibrationally excited molecule, in accord with Sekikawa et al.
We attribute this latter effect to the photochrome being formed from vibrationally excited keto molecules as illustrated in Fig. \ref{fig:jablonski}a).
This was previously proposed by Rosenfeld et al.\cite{rosenfeld_PHOTOCHROMICANILSStructPHOTOISOMERSThermRelaxProcess1973} and others,\cite{zgierski_JChemPhys2000,kownacki_ChemicalPhysicsLetters1994,mitra_PhysChemChemPhys2003,barbara_JAmChemSoc1980} but has since been disputed.\cite{ziolek_PhysChemChemPhys2004,sekikawa_JPhysChemA2013,sliwa_PhotochemPhotobiolSci2010}
This conclusion is further supported by the impact of Ar clustering on the dynamics as discussed below.

While the clustering of SA with Ar has dramatic changes on the observed \emph{dynamics} of the fluorescent state, captured by GA components B$_1$ and B$_2$, the \emph{spectrum} of the flourescent keto state is essentially unchanged.
This is not unexpected given the weak solvent dependence observed in previous TA work. \cite{mitra_PhysChemChemPhys2003,mitra_ChemicalPhysicsLetters1998}
Our justifications for assigning B$_1$ and B$_2$ to the excited keto are the same as discussed in the isolated molecule, namely correspondence with theory and previous work. 
The keto lifetime increases from 1.8 ps to a biexponential decay with $\tau_{\text{B}_1} = 32$ ps and $\tau_{\text{B}_2}$ = 1030 ps when measured with 1 bar Ar stagnation pressure.
These lifetimes continue to increase with increasing Ar carrier pressure, which we attribute to increasing the fraction of SA molecules in the beam clustered with Ar.

As mentioned before, Ar clustering can affect the dynamics in two ways: (1) dissipation of vibrational energy and (2) steric hindrance of large amplitude motions.
From the hot/cold comparison in the isolated molecule, where we observe only a small effect of vibrational energy of the dynamics, we conclude that the dominant effect is steric hindrance.
Ar atoms clustered around the central bonds can inhibit the large rotation necessary to reach CI1.
Similar overall trends were observed in SA in different environments using ps time-resolved fluorescence, with only minor changes (factors of 2-5) in lifetimes across a wide range of solvents and a few hundredfold increase in matrices and glasses.\cite{barbara_JAmChemSoc1980}

While the excited keto population giving rise to fluorescence is stabilized upon clustering with Ar, the photochrome yield we observe (captured by B$_3$), as judged by the ratio of the photochrome signal compared to to the initial TAS signal amplitudes, is the same in SA:Ar as the isolated molecule.
This strongly supports a parallel channel for the production of the photochrome, separate from the relaxed fluorescent keto state.
After ESIPT, keto geometries are formed with a wide range of internal energies. 
The B$_1$ and B$_2$ DAS components are very similar, indicating a similar electronic state and geometry, and we assign B$_1$ to the hotter portion of the keto population.
The absence of a spectral shift between B$_1$ and B$_2$, i.e. B$_2$ is not a shifted IVR product of B$_1$, further supports a parallel model.
In the proposed dynamic scheme of Fig \ref{fig:jablonski}b), the vibrationally more energetic population of keto geometries is formed promptly and reaches CI1 more easily.
This ``hotter'' population can then continue to isomerize on the ground state to the trans keto photochrome state.
Note that the initial internal energy is similar in the cold isolated and SA:Ar cases, and thus the hot keto population fraction and photochrome yields are the same, whereas the photochrome yield for the hot isolated molecule is larger than both jet-cooled SA and SA:Ar.

The vibrationally cooler population, to which we assign to B$_2$, like A$_1$, represents the state responsible for fluorescence. 
Instead of forming the photochrome, this population returns to the ground state cis keto and, eventually, undergoes back proton transfer to the S$_0$ enol minimum.
In this model, the initially excited enol is the common precursor to both channels that produce the photochrome and fluorescent state, as proposed by Zgierski and Grabowska.\cite{zgierski_JChemPhys2000} 
Note that biexponential nature and hot keto population is not observed in the isolated molecule case because the hot population can go through CI1 and form the photochrome much more quickly, and this is not resolvable with our time resolution. 
The major difference between the isolated and SA:Ar cases, highlighted in Fig.\ \ref{fig:jablonski} is the relaxation mechanism from the relaxed fluorescent state. 
In the isolated molecule, this decay to the ground state keto takes 1.8 ps and occurs predominantly via internal conversion to CI1 due to the short excited state lifetime and reduced fluorescence yield. 
In SA:Ar, the clustering shuts off the internal conversion pathway and forces radiative decay to be the dominant pathway which we observe as an increase in fluorescence and excited state lifetime. 

\section{Conclusions}

In this work we have combined cavity-enhanced transient absorption spectroscopy with AIMS/TD-CASCI calculations to study the dynamics of salicylidenaniline after excitation to S$_1$. 
This study introduces several new paradigms in ultrafast spectroscopy.
For one example, our experiments on SA in Ar clusters are analogous to matrix isolation, where ultrafast spectroscopy has previously been very difficult and limited mostly to ``action''-based methods based on fluorescence detection.\cite{engel_PhysRevB2006,guhr_JPhysChemA2002,thon_JPhysChemA2013}
Using the CE-TAS method we have shown how one can effectively record conventional ultrafast transient absorption (i.e. direct absorption) measurements in rare-gas matrix environments with rapid sample refreshment.
For another example, we have shown how long-lived TA signals with lifetime $\tau \gtrsim 1/f_{\text{rep}}$ can be recovered via careful analysis of data at negative pump/probe delays and the modified GA model of equation (\ref{eqn:CETAS_GA}).
These new techniques can be applicable for both future CE-TAS studies and other ultrafast spectroscopy contexts.

The TD-CASCI technique presented here allows for quickly simulating TAS spectra from AIMS calculations which, when compared to experimental spectra, facilitate spectral assignment. 
With gas phase TAS measurements, we are measuring the free molecular dynamics which makes direct simulation easier without the need for solvent models.

By combining these new experimental and theoretical techniques, we provide insight into the relaxation dynamics of SA, summarized in Fig. \ref{fig:jablonski}. 
After prompt ESIPT, keto geometries are formed with a broad energy distribution, the hottest of which rapidly internally convert through CI1 and isomerize to the trans-keto photochromic state. 
The colder keto population relaxes more slowly through CI1 to reach the cis-keto ground state in 1.8 ps for the isolated molecule and eventually return to the enol S$_0$ minimum.
Increasing the internal energy by lowering the gas pressure results in more hot keto population and more yield of the long-lived photochrome.
Embedding the SA molecule in Ar clusters sterically hinders isomerization such that only the hot keto population can undergo internal conversion to the photochrome state, with a dramatic increase in the S$_1$ lifetime recorded in TAS and also the fluorescence yield.
We find no experimental evidence for the proposed secondary enol twist relaxation channel, but our calculations indicate that it likely lies outside our detection window.

\begin{acknowledgments}
This work was supported by American Chemical Society Petroleum Research Fund under grant number 62125-ND6, the U.S. National Science Foundation under award number 2102319 and the U.S. Air Force Office of Scientific Research under grant number FA9550-20-1-0259.
GK acknowledges support from the European Union's Horizon 2020 research and innovation programme under the Marie Sklodowska-Curie Grant Agreement No 101028278. 
AM and BGL acknowledge the Institute for Advanced Computational Science and Stony Brook University for funding.
\end{acknowledgments}

\section*{AUTHOR DECLARATIONS}

\subsection*{Conflict of Interest}
The authors have no conflicts to disclose

\section*{Data Availability}
The data that support the findings of this study are available from the corresponding author upon reasonable request.

\bibliography{SA}

\end{document}


\title{Supplementary material: \\ Ultrafast internal conversion and photochromism in gas-phase salicylideneaniline}
	\author{Myles C. Silfies}
	\affiliation{Department of Physics and Astronomy, Stony Brook University, Stony Brook, New York 11794, USA}
	\author{Arshad Mehmood}
	\affiliation{Department of Chemistry, Stony Brook University, Stony Brook, New York 11794, USA}
	\affiliation{Institute for Advanced Computational Science, Stony Brook University, Stony Brook, New York 11794, USA}
	\author{Grzegorz Kowzan}
	\affiliation{Department of Chemistry, Stony Brook University, Stony Brook, New York 11794, USA}
	\affiliation{Institute of Physics, Faculty of Physics, Astronomy and Informatics, Nicolaus Copernicus University in Toru\'{n}, ul. Grudziadzka 5, 87-100 Toru\'{n}, Poland}
	\author{Edward G. Hohenstein}
	\affiliation{QC Ware Corporation, Palo Alto, California 94301, USA}
	\author{Benjamin G. Levine}
	\affiliation{Department of Chemistry, Stony Brook University, Stony Brook, New York 11794, USA}
	\affiliation{Institute for Advanced Computational Science, Stony Brook University, Stony Brook, New York 11794, USA}
	\author{Thomas K. Allison}
	\affiliation{Department of Physics and Astronomy, Stony Brook University, Stony Brook, New York 11794, USA}
	\affiliation{Department of Chemistry, Stony Brook University, Stony Brook, New York 11794, USA}
	\email{thomas.allison@stonybrook.edu}
{
\let\clearpage\relax
\maketitle
}
\tableofcontents

\section{Time-resolved signal and global fitting routine}
\label{sup:sec:fit}
As discussed in detail previously,\cite{silfies_PhysChemChemPhys2021} the pump-probe signal measured in the cavity-enhanced spectrometer includes a subtracted "reference" signal which is recorded from a copy of the probe comb coupled into the probe cavity backwards and delayed by $\approx 5 ns$.
The reference signal is subtracted from the probe signal in an autobalanced detector, resulting in a recorded molecular signal, $\Delta S(\lambda,t)$ of the form
\begin{equation}
	\Delta S(\lambda,t) = \beta(\lambda)[\Delta I(\lambda,t) - \Delta I(\lambda,t + 5 \textrm{ ns})]  
	\label{eqn:DeltaS}
\end{equation} 
Where $\beta = \frac{\pi}{\mathcal{F(\lambda)}}\frac{1}{I_{probe}}$ is the inverse of the cavity enhancement factor divided by the static probe intensity (which matches the reference due to autobalancing).
$\Delta I$ is the pump-induced change in absorption and $\tau$ is the pump/probe delay. 
The reference subtraction reduces common-mode noise which increases absorption sensitivity and also removes quasi-static signal offset due to long-lived molecular dynamics, or multi-excitation effects caused by the necessarily high repetition rate of the experiment (100 MHz).
Cavity finesse, $\mathcal{F}$, is measured at every probe wavelength immediately following pump/probe scans to use in signal calculations.
\begin{figure}
	\includegraphics[width=\linewidth]{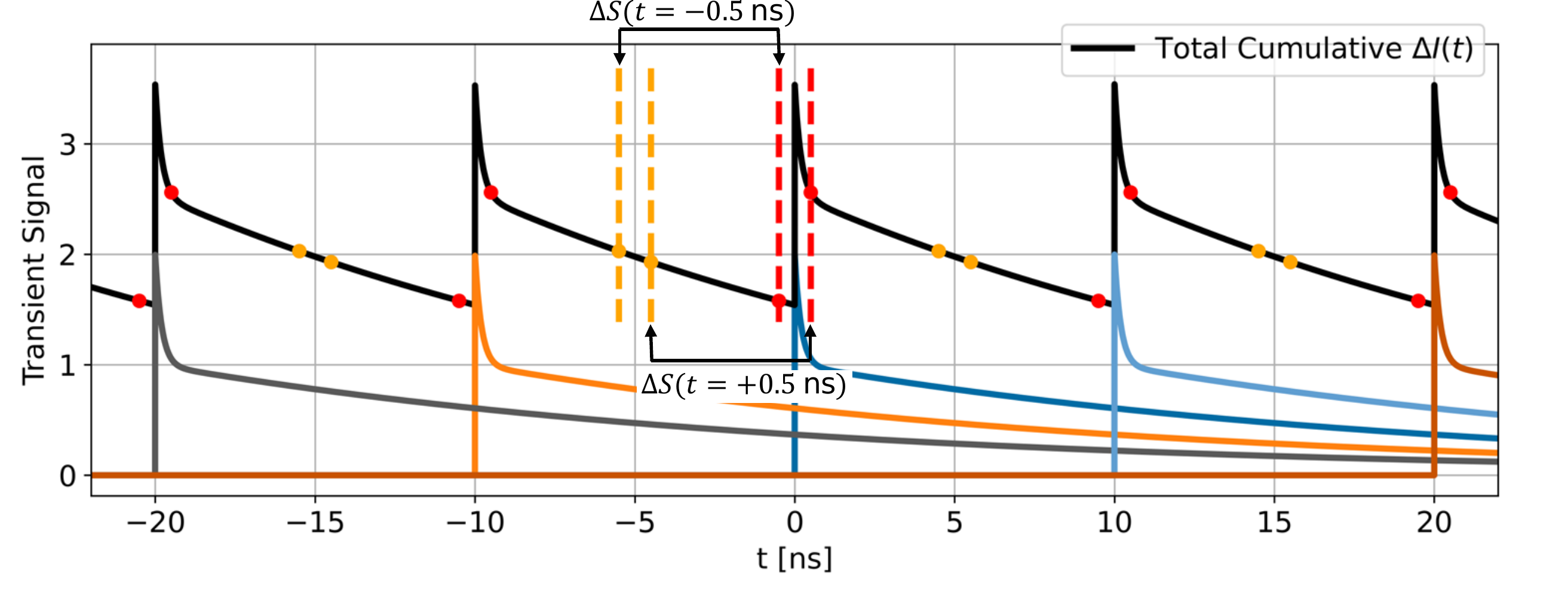}
	\caption{Illustration of origin of negative delay signal. The sample is pumped at $f_{\text{rep}} = 100$ MHz. Each excitation pulse launches a bi-exponential decay molecular response (colored lines) with $\tau_1 = 0.2$ ns and $\tau_2 = 20$ ns = $2/f_{\text{rep}}$. The CE-TAS signal $\Delta S(\tau)$ is constructed from the difference between the total cumulative steady state change in absorbance signal $\Delta I$ sampled at the probe pulses (red dots) and the reference pulses (orange dots) via equation (\ref{eqn:DeltaS}). For molecules with signals with lifetimes on the order of $1/f_{\text{rep}}$ or longer, this can lead to nonzero signals before time zero, as illustrated in Fig. \ref{fig:dS}.}
	\label{fig:dIss}
\end{figure}

\begin{figure}
	\includegraphics[width=.6\linewidth]{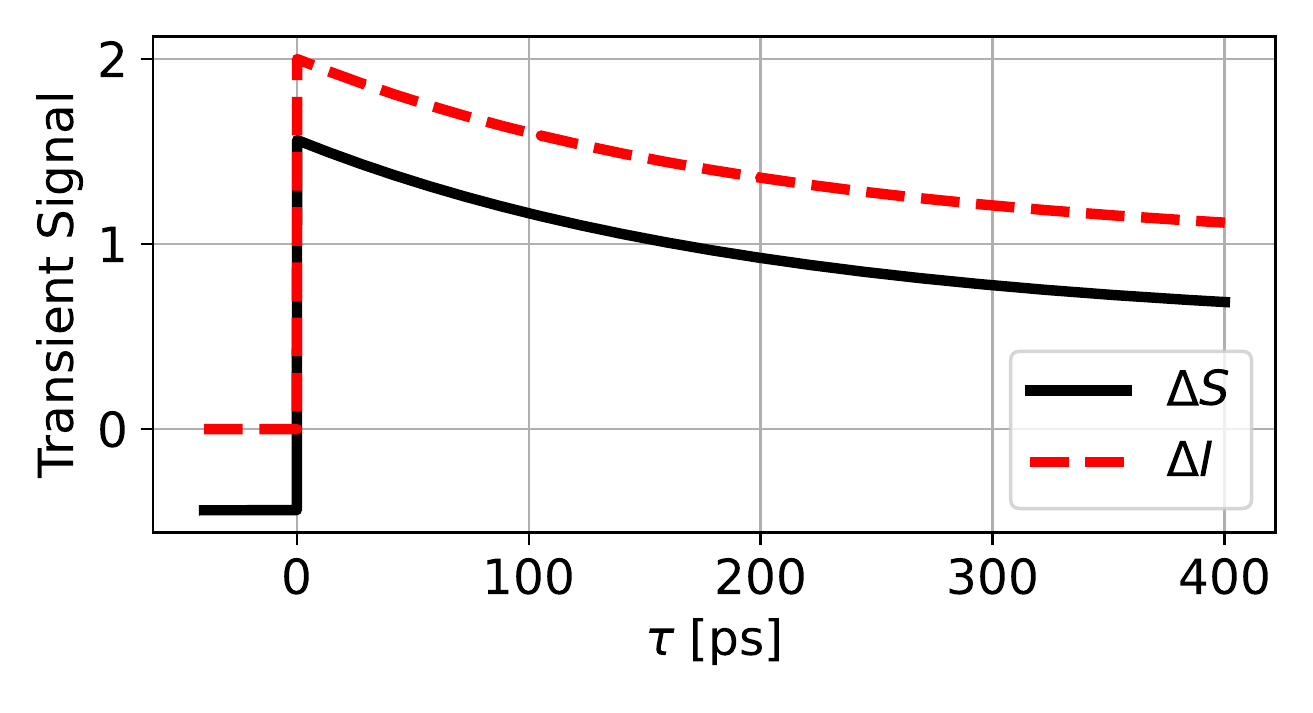}
	\caption{CE-TAS signal $\Delta S$ corresponding the scenario of Fig. \ref{fig:dIss}. A sizable negative signal is seen before time zero.}
	\label{fig:dS}
\end{figure}

Fig. \ref{fig:dIss} shows how $\Delta I$ is recorded in CE-TAS for an example system featuring biexponential decay.
The multicolored scans separated by 10 ns (1/f$_\textrm{rep}$) represent individual molecular responses from each subsequent pump excitation and the black line is the sum of these which would be recorded by the actual spectrometer.
To construct $\Delta S(t)$ from $\Delta I(t)$, the reference signal, $\Delta I(t+5ns)$  is subtracted as shown for 2 example points by the vertical dashed lines in Fig. \ref{fig:dIss}.
As seen in the figure, the point at -0.5 ns captures an inverted molecular signal from the previous excitations, resulting in a nonzero offset at negative delays.
The resulting $\Delta S(t)$ is shown in Fig. \ref{fig:dS} as well as the original $\Delta I(t)$.
It is clear that the short time dynamics are identical on both and only the long-lived signal displaces $\Delta S(t)$ vertically.

The TA spectra, both experimental and simulated, are fit to a sum of exponential decays convolved with the wavelength-dependent instrument response.
\begin{equation}
	G(\lambda,t) = \textrm{IRF}(\lambda,t) \otimes \sum_\textrm{n} \textrm{X}_\textrm{n}(\lambda) \exp (-t /\tau_\textrm{n})
	\label{eqn:GA}
\end{equation}
Where $A_n(\lambda)$ is the decay associated spectra for each time constant $\tau_\textrm{n}$.
To model the full experimental $\Delta S(t)$, including the pre-time 0 ``artifact'' discussed above, we directly simulate the reference subtraction and sum over the next 20 probe-reference pulses by shifting the time axis of the fit by the time spacing between pulses (1/f$_\textrm{rep}$):
\begin{equation}
	\Delta S_{\text{model}} (\lambda,t) = \sum_{m=0}^{N} G(\lambda,t+m/f_{\text{rep}})-G(\lambda,t+ m/f_{\text{rep}} + 5 \textrm{ ns}) 
	\label{eqn:S_model}
\end{equation}
N = 20 pulses was chosen as the upper limit of the sum based on the average transit time of the molecular beam across the diameter of the probe beam.
The difference between the fit and experimental/simulated data is minimized using a Levenberg–Marquardt fit algorithm. 
For all experimental spectra, the instrument response is derived from a fit of the signal rising edge to an error function at each probe wavelength. 
This approximation was used since previous measurements of SA concluded that the proton transfer occurred below our instrument time resolution \cite[]{sekikawa_JPhysChemA2013} so a step-like molecular signal could be assumed. 
The fitting routine was run multiple (>50) times starting from different points in parameter space and the global minimum was used as the final optimum fit.
\begin{figure}
	\includegraphics[width=.6\linewidth]{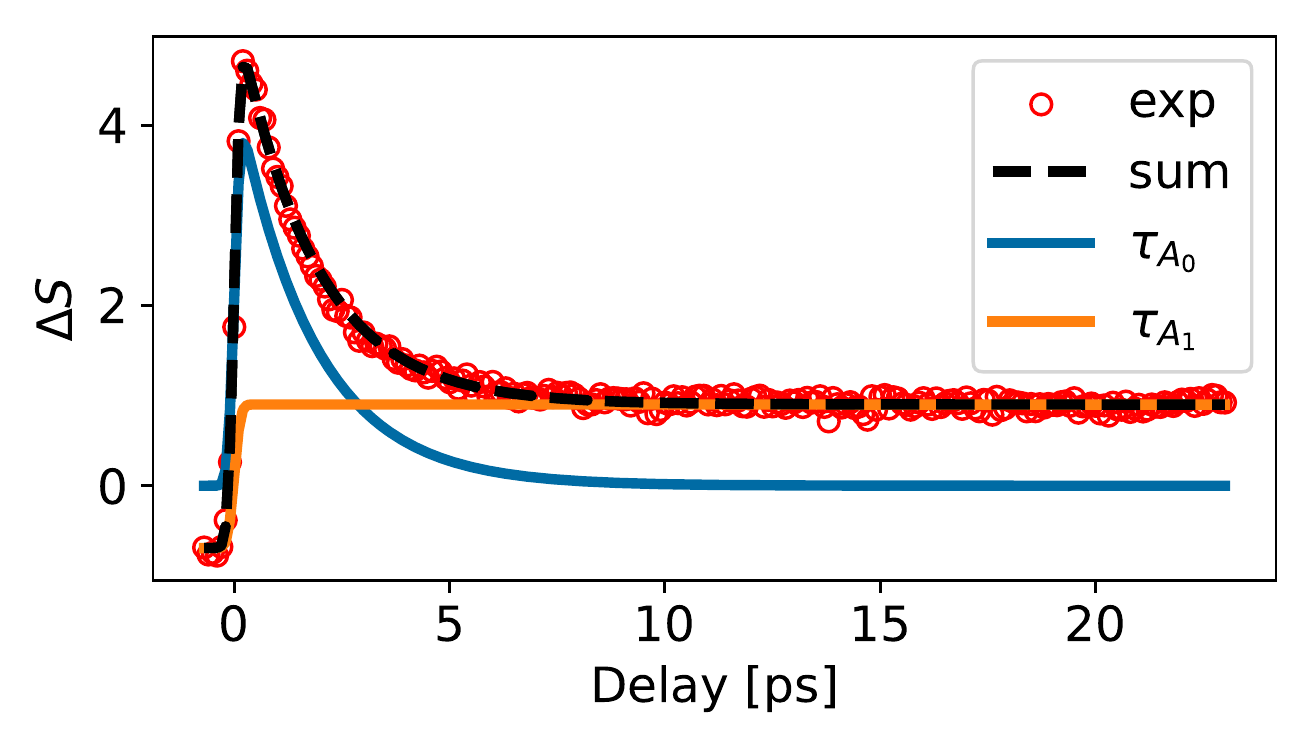}
	\caption{Single scan fit at 455 nm with isolated time constant contributions using equation \ref{eqn:S_model_flip}.}
	\label{fig:comp_lineouts}
\end{figure}

By combining equations \ref{eqn:GA} and \ref{eqn:S_model} and reversing the order of summations, the fit model can also be written as:
\begin{equation}
	\Delta S_{\text{model}} (\lambda,t) = \textrm{IRF}(\lambda,t) \times \sum_\textrm{n}\textrm{X}_\textrm{n}(\lambda) \times \sum_{m=0}^{20} \left[\exp(-(t+m/\textrm{f}_\textrm{rep})/\tau_\textrm{n})-\exp(-(t+m/\textrm{f}_\textrm{rep}+5 \textrm{ns})/\tau_\textrm{n})\right]
	\label{eqn:S_model_flip}
\end{equation}
where the different time components can be isolated. 
An example of this procedure is shown in Fig. \ref{fig:comp_lineouts} for a single scan of isolated SA.
The long-lived signal captures all of the pre-time 0 signal and the short time component is unaffected by the multi-pulse fitting.

\subsection{Goodness of fit}
To verify the applicability of the fit models used in the main text, fits using the same approach outlined above for different number of fit components are shown here.

\begin{figure}
	\includegraphics[width=.6\linewidth]{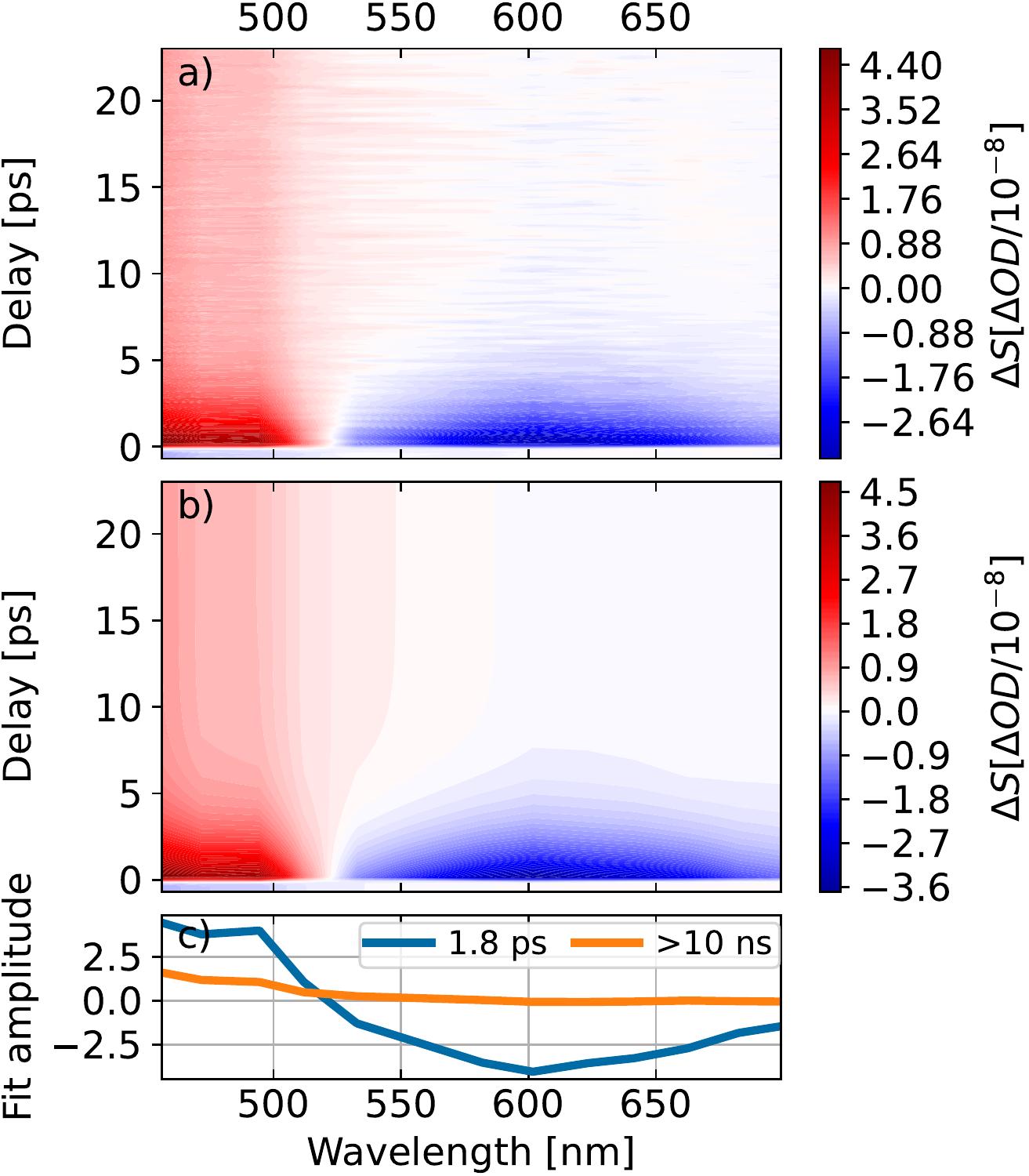}
	\caption{2 component parallel fit model on isolated SA. a) Experimental spectrum. b) Spectrum returned from fit. c) DAS from b).}
	\label{fig:He_ds_2}
\end{figure}

\begin{figure}
	\includegraphics[width=.6\linewidth]{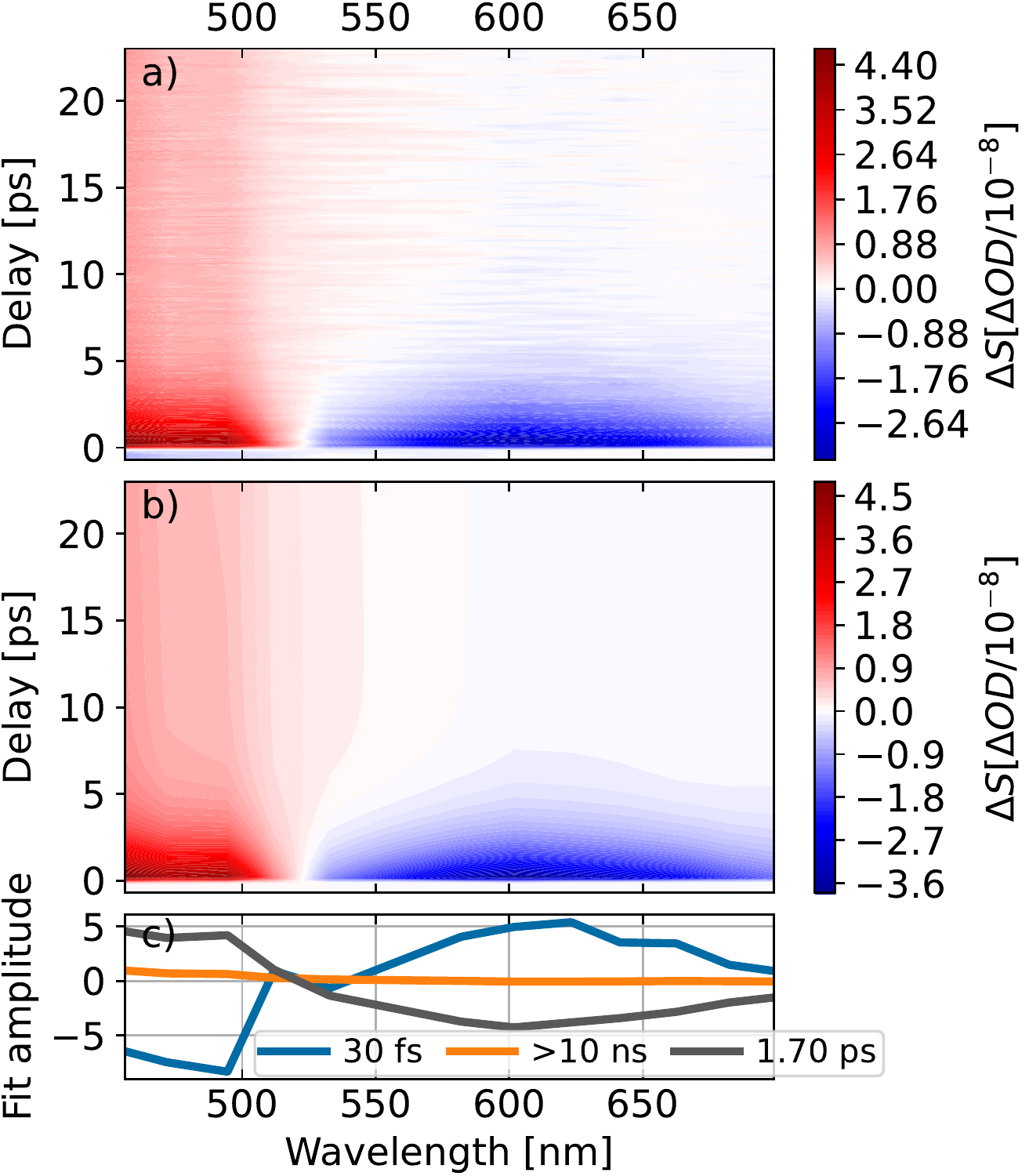}
	\caption{3 component parallel fit model on isolated SA. a) Experimental spectrum. b) Spectrum returned from fit. c) DAS from b).}
	\label{fig:He_ds_3}
\end{figure}

\begin{figure}
	\includegraphics[width=.6\linewidth]{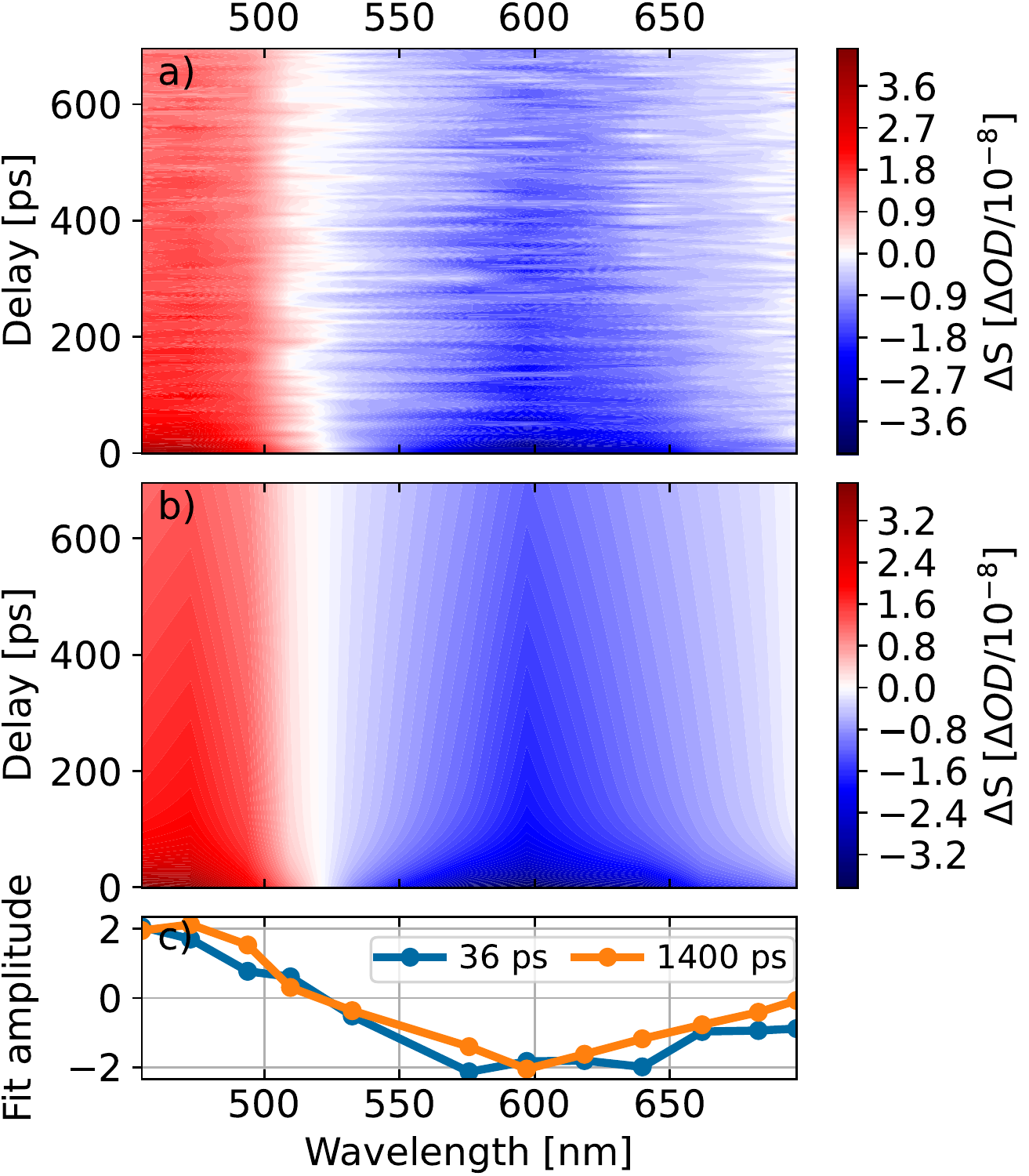}
	\caption{2 component parallel fit model on SA:Ar. a) Experimental spectrum. b) Spectrum returned from fit. c) DAS from b).}
	\label{fig:Ar_ds_2}
\end{figure}

\begin{figure}
	\includegraphics[width=.6\linewidth]{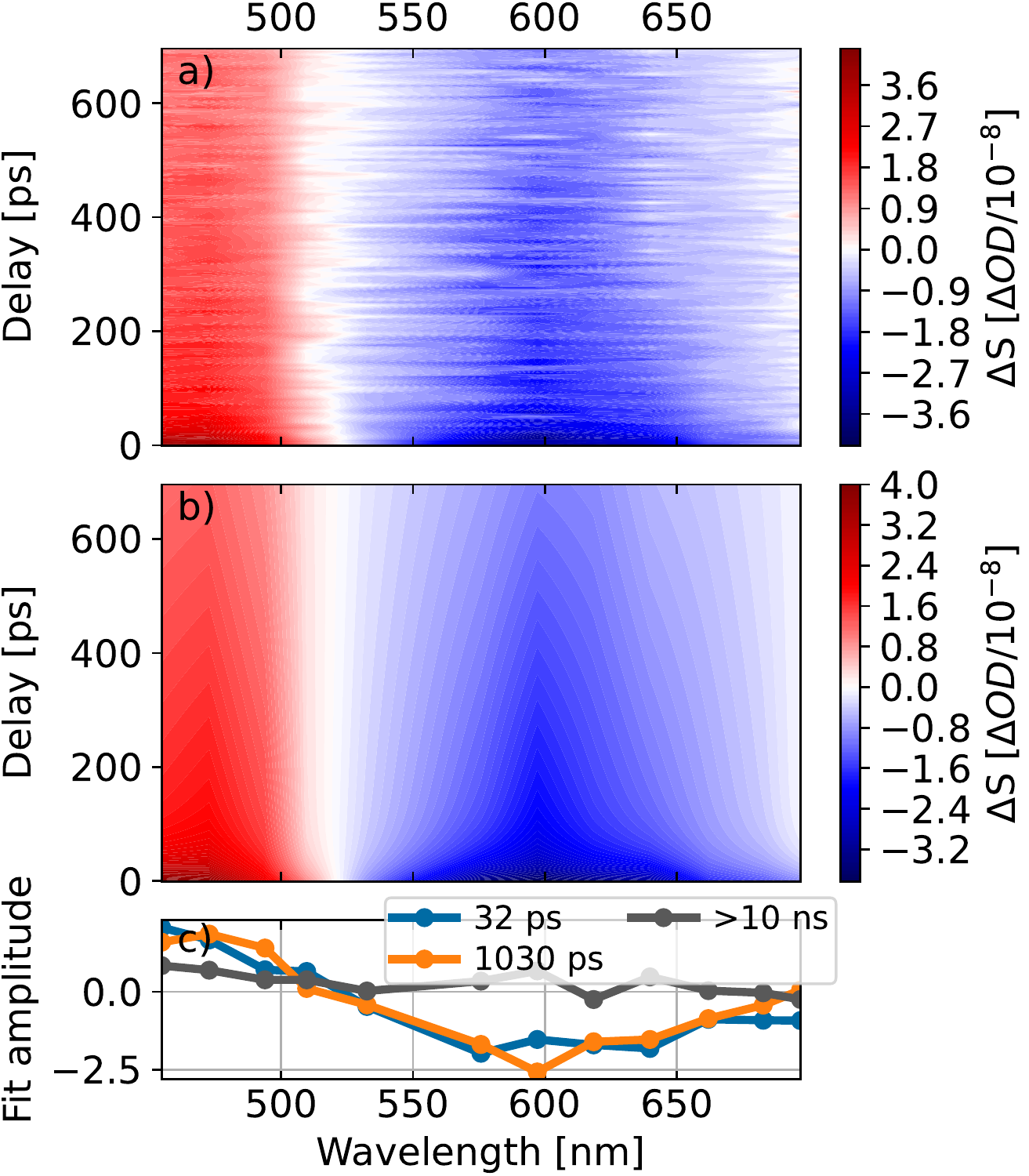}
	\caption{3 component parallel fit model on SA:Ar. a) Experimental spectrum. b) Spectrum returned from fit. c) DAS from b).}
	\label{fig:Ar_ds_3}
\end{figure}

\begin{figure}
	\includegraphics[width=.6\linewidth]{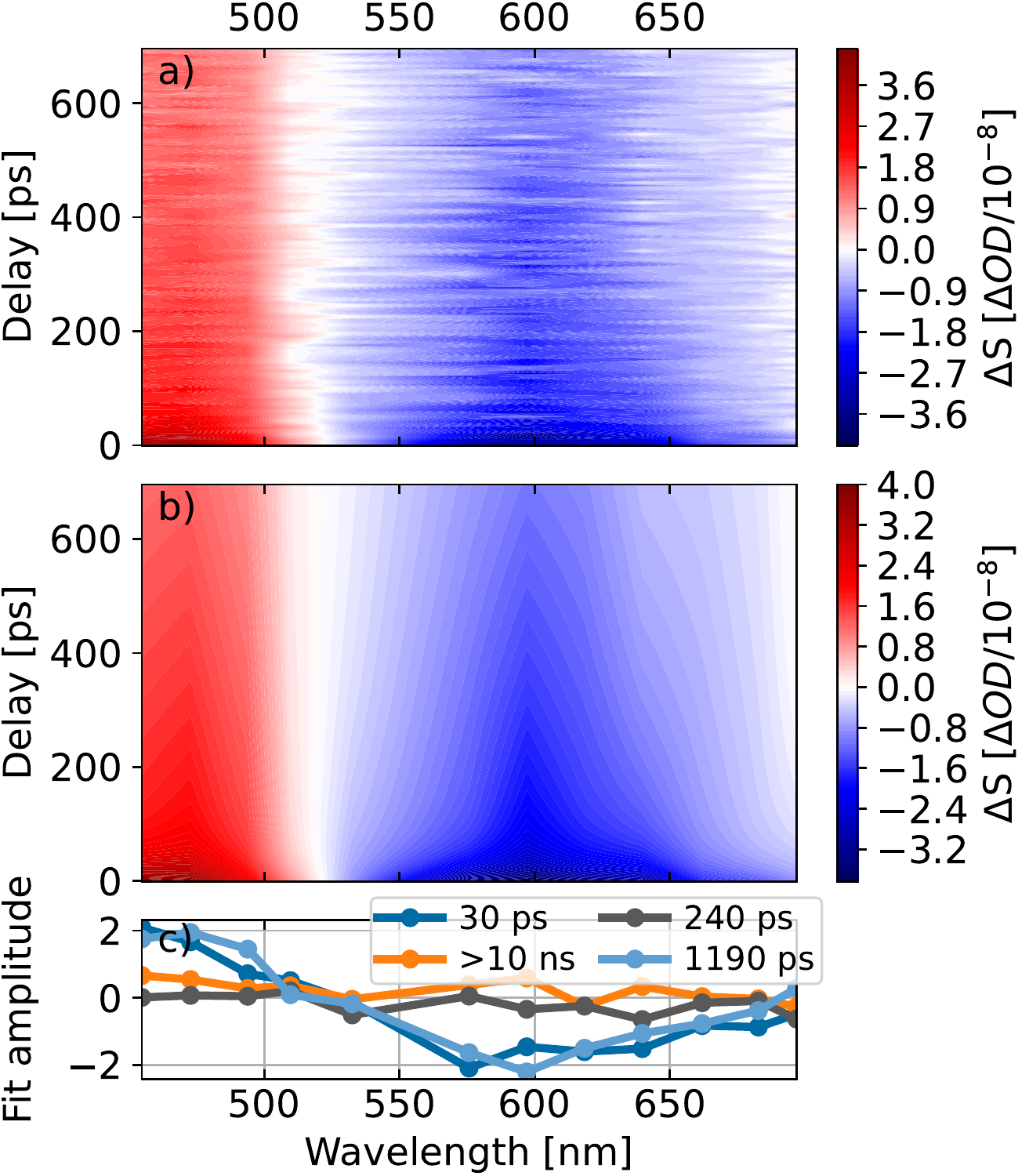}
	\caption{4 component parallel fit model on SA:Ar. a) Experimental spectrum. b) Spectrum returned from fit. c) DAS from b).}
	\label{fig:Ar_ds_4}
\end{figure}

\begin{figure}
	\includegraphics[width=.9\linewidth]{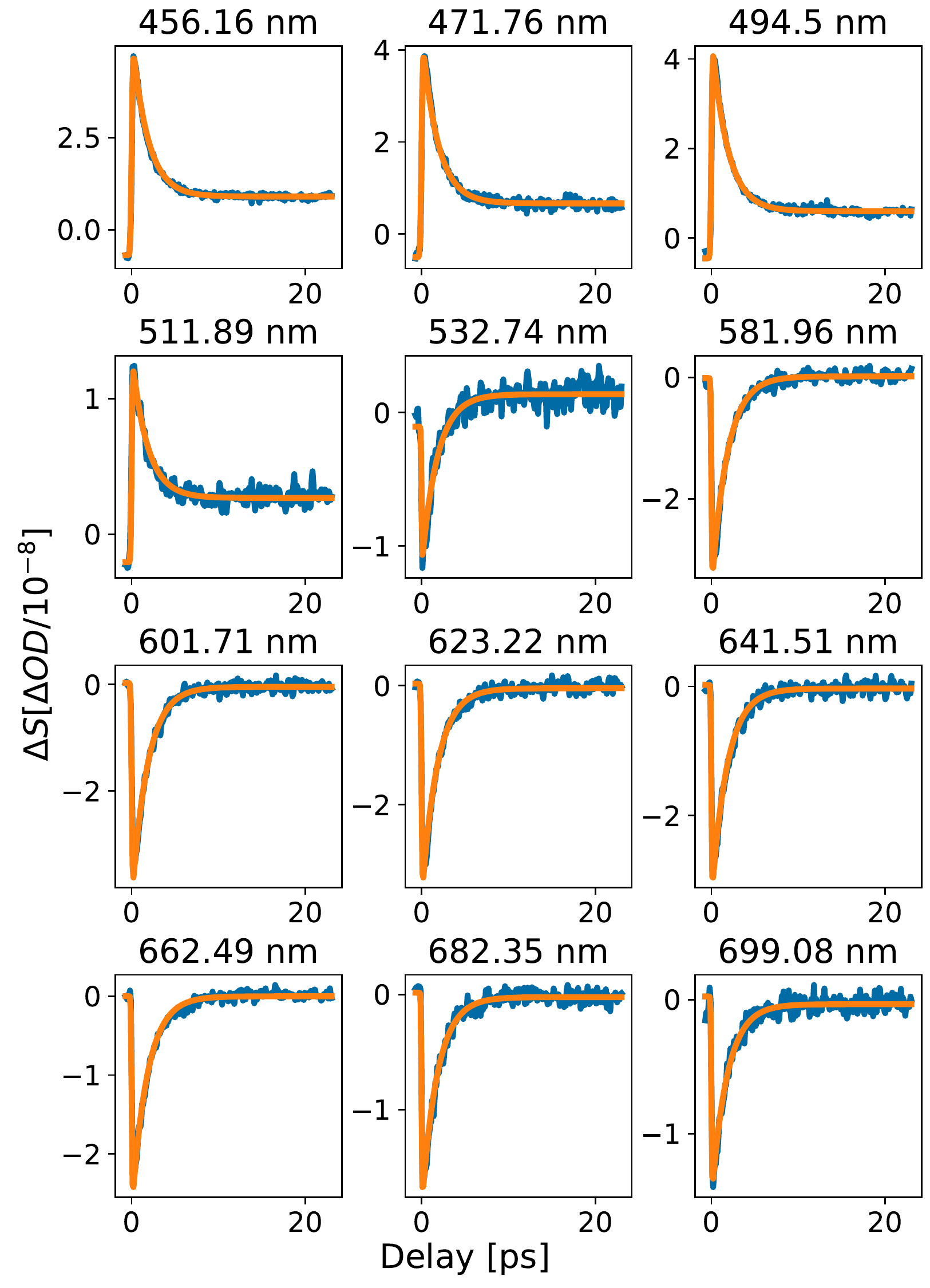}
	\caption{Isolated SA spectral lineouts including corresponding 2 component fit traces.}
	\label{fig:He_all}
\end{figure}

\begin{figure}
	\includegraphics[width=.9\linewidth]{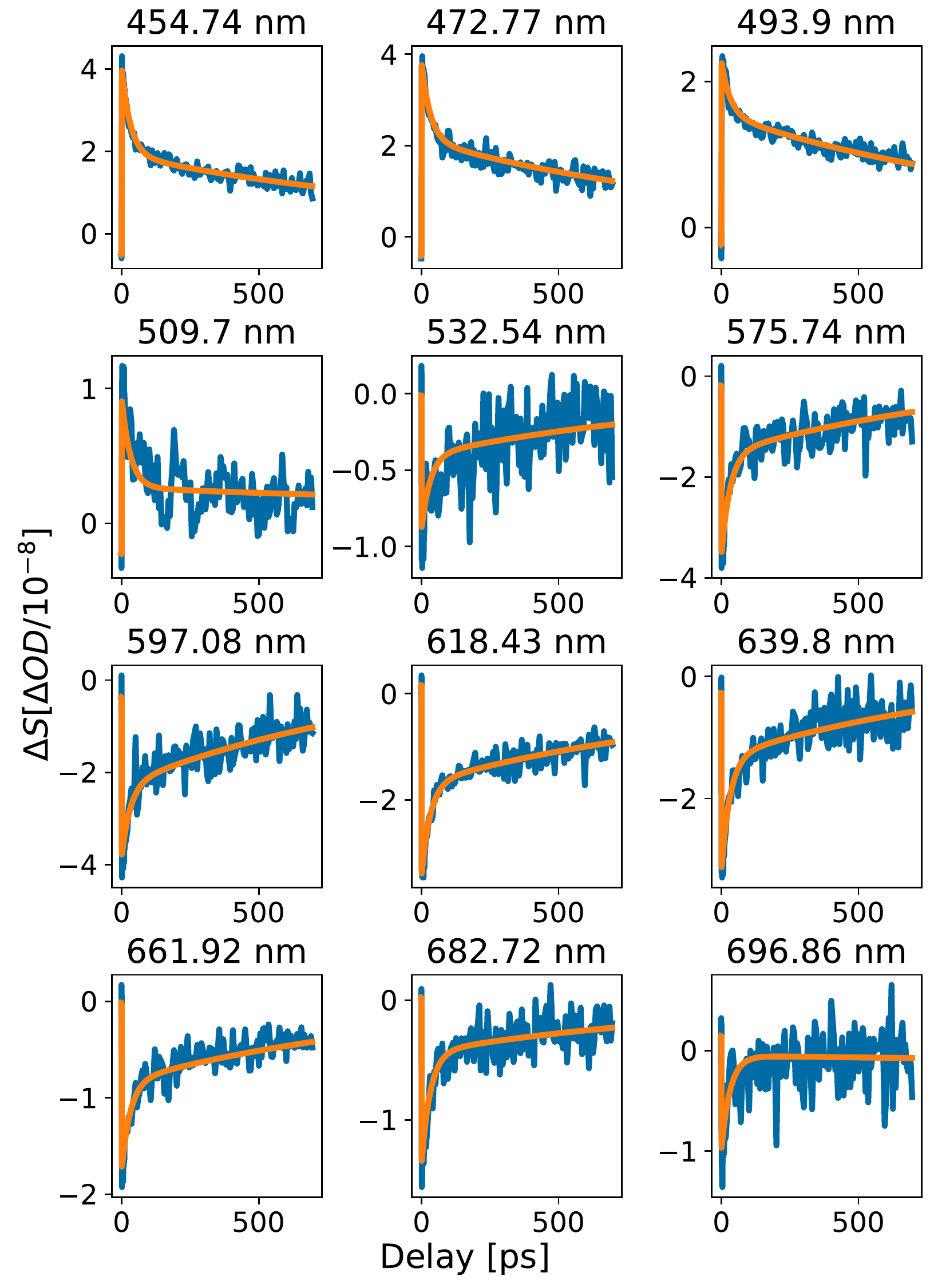}
	\caption{SA:Ar spectral lineouts including corresponding 3 component fit traces.}
	\label{fig:Ar_all}
\end{figure}

\section{Instrument response function}
\begin{figure}
	\includegraphics[width=.6\linewidth]{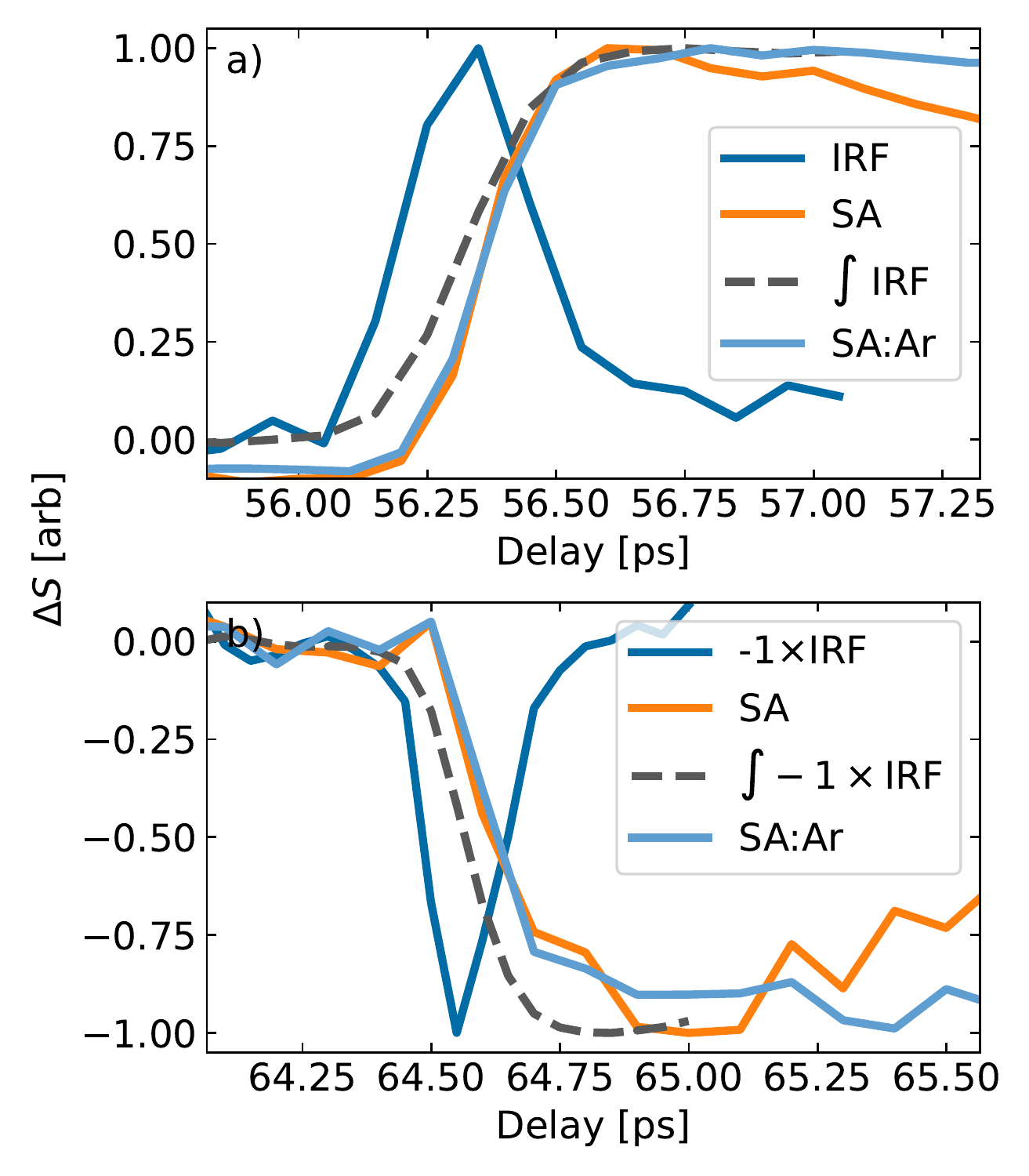}
	\caption{Comparison of the independent IRF and the isolated SA and SA:Ar signals at a) 462 nm and b) 616 nm.}
	\label{fig:int_IRF}
\end{figure}

To independently verify the instrument response function (IRF) of the spectrometer, we measure the two-photon absorption of carbon disulfide at several wavelengths immediately following SA measurements.
The results are shown in Fig. \ref{fig:int_IRF}. 
Also shown is the cumulative integral of the carbon disulfide data to better compare to the error function response of SA.
We do not observe a meaningful delay in time SA signal onset, any apparent shift is within the experimental and/or fit error. 

\section{Computational Methods}
\subsection{Post-processing of Non-adiabatic Molecular Dynamics Trajectories}
The gas-phase TAS of SA was simulated by post-processing the trajectories of 2 ps excited-state non-adiabatic molecular dynamics (NAMD) simulations.
The trajectories of \textit{Ab-initio} Multiple Spawning (AIMS) \cite[]{ben-nun_JPhysChemA2000,ben-nun_ChemicalPhysics2000,curchod_ChemRev2018} NAMD simulations were thankfully provided by Pijeau \textit{et al.} and the details of the simulation are described in reference \onlinecite{pijeau_JPhysChemA2018}.
The AIMS data consisted of more than 2000 trajectory basis functions distributed between the S\textsubscript{1} and S\textsubscript{0} electronic states and contained a total of 854,760 geometric conformations with 715,618 conformations belonging to the S\textsubscript{1} electronic state.
Both the S\textsubscript{1} and S\textsubscript{0} geometries were time-resolved by using 1000 a.u ($\sim$24.2 fs) time intervals for a total of 84 sampling windows.
Within each window, 80 distinct conformations were selected by using the \textit{k}-means clustering algorithm weighted by the electronic population of the trajectory basis functions representing the conformation.
The weighted \textit{k}-means clustering used the pairwise RMSD distance matrices which were calculated by using the MDTraj \cite{mcgibbon_BiophysJ2015} library.
Within each sampling window, the cluster centroid was translated and rotated by using the Kabsch algorithm \cite{kabsch_ActaCrystA1976} to minimize the RMSD between the centroid and optimized S\textsubscript{0} enol geometry oriented along the \textit{z}-axis of its transition dipole.
A total of 6720 representative time-resolved conformations on S\textsubscript{1} were selected for the TAS simulations.                 
Because simulations were stopped upon quenching of the majority of the population to S\textsubscript{0}, we did not include S\textsubscript{0} population in the TAS calculation.

\subsection{Simulation of TAS}
\begin{figure}
	\includegraphics[width=.6\linewidth]{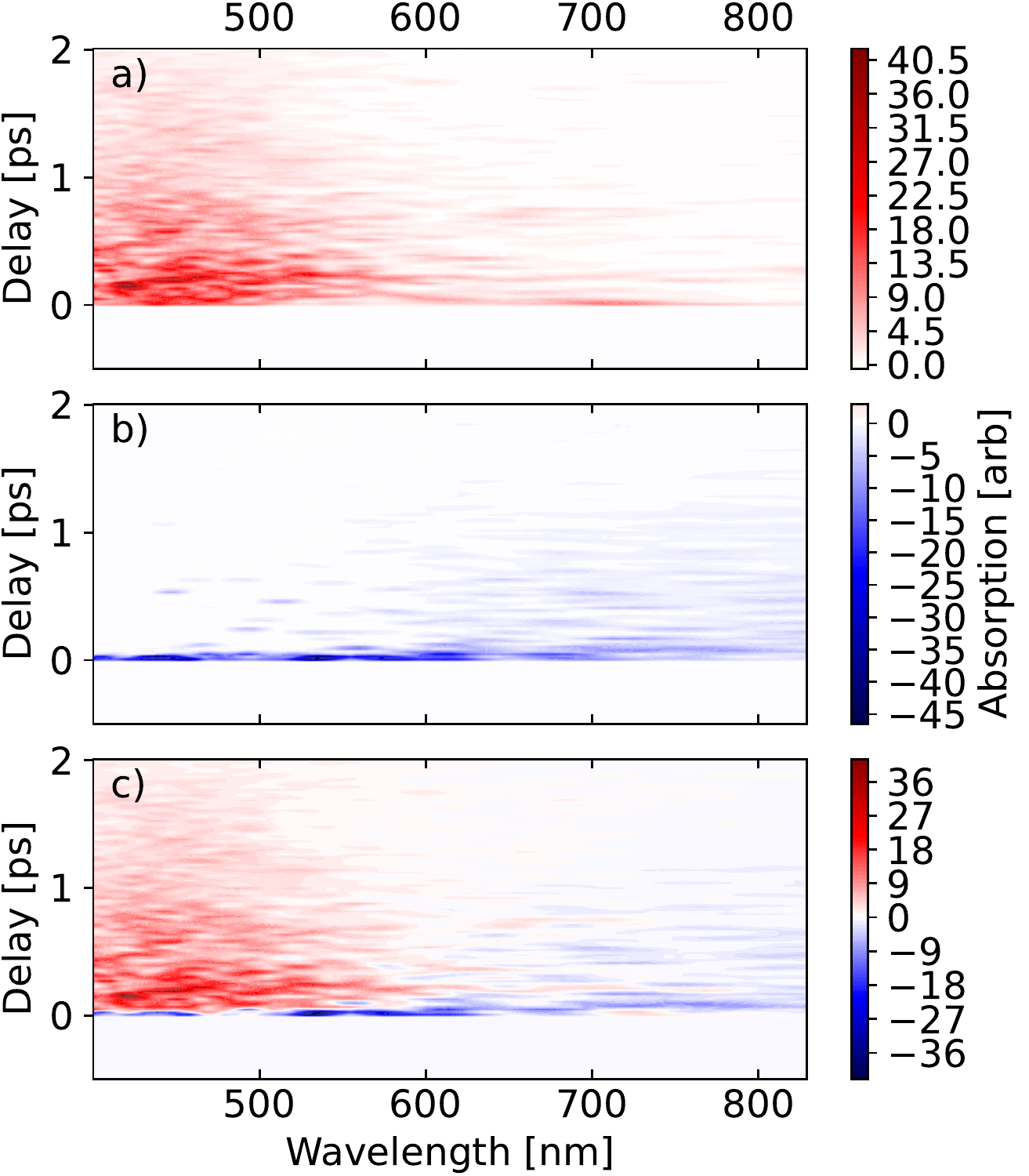}
	\caption{Raw, unconvolved simulated spectrum. a) ESA b) SE c) total signal.}
	\label{fig:noconv}
\end{figure}
\label{sup:sec:sim}
The simulation of TAS uses our GPU accelerated implementation of the Time-Dependent Complete Active Space Configuration Interaction (TD-CASCI) \cite{peng_JChemTheoryComput2018} method, implemented in the TeraChem suite of programs \cite{seritan_WIREsComputMolSci2021}.
The details of the method can be found in Ref. \onlinecite{peng_JChemTheoryComput2018}.
Briefly, in TD-CASCI the time-dependent electronic wave function $\Psi(t)$, is expressed as a linear combination of Slater determinants, $\{\Phi_{K}\}$,
\begin{equation}
    \Psi(t) = \sum_{K} C_{K}(t) \Phi_{K}
\end{equation}
where $C_{K}(t)$ are the time-dependent  CI expansion coefficients. 
The time-dependent wave function is propagated by numerically solving the time-dependent Schrodinger equation, 
\begin{equation}
    i\frac{\partial \mathbf{C}(t)}{\partial t}=\mathbf{H}(t) \mathbf{C}(t)
\end{equation}
Our GPU-accelerated implementation calculates the $\mathbf{H}\mathbf{C}$ products \textit{on-the-fly}, avoiding the computationally expensive procedures of  building, storing, and/or diagonalizing the CI Hamiltonian.
Propagation is carried out via a second-order symplectic split operator integrator.

To include the explicit field effects, the Hamiltonian at any time \textit{t} is expressed in the electric dipole approximation, 

\begin{equation}
    \hat{H}(t)=\hat{H}_{0}-\hat{\boldsymbol{\mu}} \cdot \mathbf{d} E(t)    
\end{equation} 
where $\hat{H}_{0}$ is the field-free molecular Hamiltonian, and $\hat{\boldsymbol{\mu}}$ is the molecular dipole operator.
The scalar function, $E(t)$, is the time-dependent external field strength, and $d$ is the unit vector in the field polarization direction.
The obtained time correlation function can be used to get the energy spectrum of the electronic wave function after excitation with a pulse,
\begin{equation}
    R(t)=\mathbf{C}(\varepsilon)^{\dagger} \mathbf{C}(\varepsilon+t)
\end{equation}
where $\varepsilon$ corresponds to the time at the end of the pulse.
The absorption spectrum of any electronic state can be obtained by the Fourier transform of the $R(t)$ after excitation with a $\delta$-function pulse.   
To reduce the effects of the spectral leakage, the Hanning Windowing function was applied to raw $R(t)$ prior to the Fast Fourier transform of the obtained time correlation function of Eq. 4.

The spectrum is shifted such that the zero of energy is at the initial (S\textsubscript{1}) state energy. 
Signal with positive energy corresponds to excite state absorption (ESA), while signal with negative energy corresponds to stimulated emission (SE).
The signal corresponding to the negative energy axis after shifting represents the SE.
The total spectrum, including both ESA and SE and defined only for $E\ge0$, is computed according to:

\begin{align}
R_{ESA-SE}(E) &= R(E) - R(-E)
\end{align}

Prior to the above summation, shifts are applied to the TD-CASCI spectra to provide more accurate energetics.  The ESA signal in the simulated TAS was blue-shifted by 0.944 eV which represents the difference in the $\textrm{S}_3 \leftarrow \textrm{S}_1$ vertical excitation energy of the S$_0$ enol ground state minimum at SA-4-FOMO(0.25)-CAS(8,8)CI and the more accurate SA-4-CAS(8,8)PT2 levels.
The $\textrm{S}_3 \leftarrow \textrm{S}_1$ transition was used as a reference due to significantly higher values of the oscillator strength  compared to the $\textrm{S}_2 \leftarrow \textrm{S}_1$ transition at the SA-4-CAS(8,8)PT2 level.
The SE signal was red-shifted by 1.595 eV, setting the S\textsubscript{0}/S\textsubscript{1} vertical excitation energy of the S\textsubscript{0} enol conformation computed at SA-4-FOMO(0.25)-CAS(8,8)CI level to the position of experimental absorption maximum \cite{mitra_PhysChemChemPhys2003}.  
The overlap of the SE and ESA and, therefore, the overall spectral shape and zero crossing is very dependent on the relative shifts applied to each signal but the overall time dynamics are independent of shifting as can be seen in Fig.\ \ref{fig:noconv}.

In order to construct the experimentally isotropic observable from a static aligned molecule, first the magnitude of the signal is constructed $R_{total}(E)=R_x(E)+R_y(E)+R_z(E)$.
The parallel, perpendicular, and magic angle signals can then be calculated from the angular dependent nonlinear absorption equation from ref \onlinecite{hochstrasser_ChemicalPhysics2001} simplified to each case:
\begin{align}
    R_{\parallel}(E) &= \frac{1}{15}(R_x(E)+R_y(E)+3R_z(E)) \\
    R_{{\perp}}(E) &= \frac{1}{15}(2R_x(E)+2R_y(E)+R_z(E)) \\
    R_{\text{MA}}(E) &= R_{total}(E)/9
\end{align}

The ESA and SE are shown in Fig. \ref{fig:noconv}a) and b).
The total, unconvolved TAS signal is shown in Fig. \ref{fig:noconv}c).
The spectrum was constructed by using the electronic spectrum of the time-resolved geometries of SA at FOMO(0.25)-TD-CAS(8,8)CI/6-31g** \cite{slavicek_JChemPhys2010} level using a $\delta$-kick with a field strength of 10\textsuperscript{24} W/m\textsuperscript{2} polarized separately along the \textit{x}, \textit{y} and \textit{z} directions of the molecular axis for a duration of 0.062 a.u (0.0015 fs).
The electronic dynamics were propagated for 15000 time steps with a step size of $\Delta$t = 0.124 a.u. (0.003 fs) which represents the $\sim$1860 a.u (45 fs) electronic dynamics.

\subsection{Twisted Enol spectrum}
\begin{figure}
	\includegraphics[width=.6\linewidth]{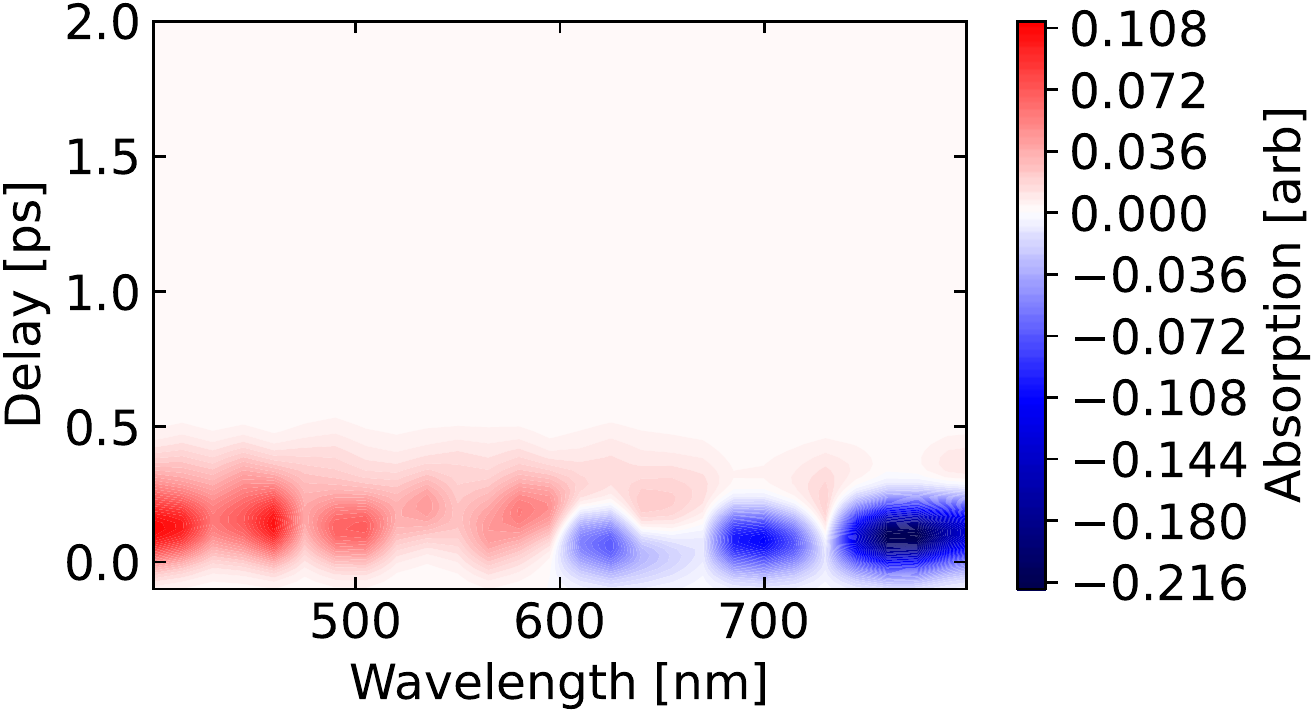}
	\caption{Theoretical TAS from enol geometries with central CN twist angles of $< 140^\circ$, indicative of the CN twist relaxation mechanism via CI2. Note that the maximum amplitude shown here is more than an order of magnitude smaller than the main spectral features on the full spectrum in the main text.}
	\label{fig:twist}
\end{figure}

The simulated TAS spectrum of twisted enol geometries is shown in Fig. \ref{fig:twist}.
As discussed in the main text, this spectrum only includes contribution from enol geometries with a CN twist angle $< 140^\circ$, indicative of internal conversion from planar geometries ($180^\circ$) towards CI2 ($\approx 90^\circ$).
Note that the maximum amplitude shown on Fig. \ref{fig:twist} is approximately one order of magnitude smaller than that shown on the full spectrum (main text Fig.\ 7a))

\subsection{Photochrome absorption simulation}
\begin{figure}
	\includegraphics[width=.6\linewidth]{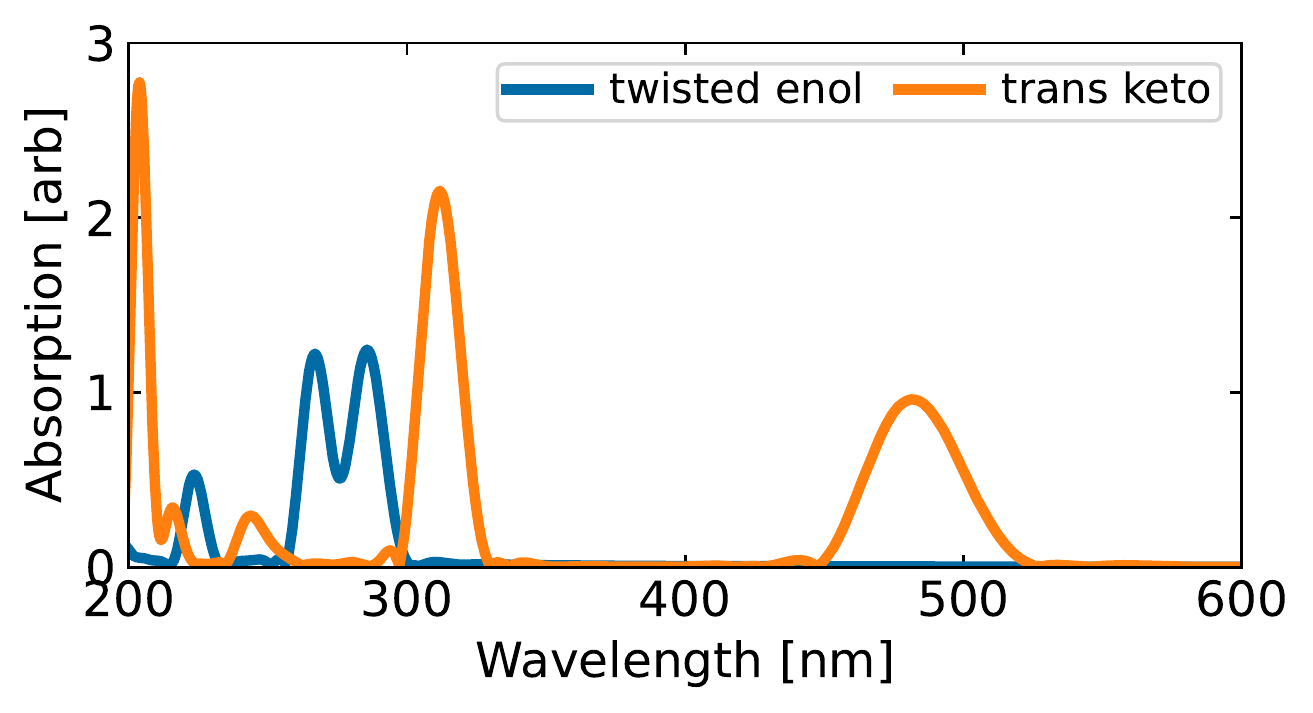}
		\caption{Ground-state absorption from twisted enol and keto photochrome candidates using optimized geometries from \onlinecite{ortiz-sanchez_JChemPhys2008}.}
	\label{fig:ortiz_abs}
\end{figure}

The simulated ground-state absorption of the two potential photochrome candidates -- the trans-keto and rotated enol -- are shown in Fig. \ref{fig:ortiz_abs}.
Optimized geometries for both conformers are from \onlinecite{ortiz-sanchez_JChemPhys2008}.
Spectra were generated using TD-CASCI method with the same level of theory and energy shifts as the SE discussed above.
\bibliography{SA_supp}